%% file: main.tex
\documentclass[12pt,a4paper]{article}

\usepackage{babel}
\usepackage[utf8]{inputenc}     
\usepackage[dvips]{graphicx}    
\usepackage{a4wide}             
\usepackage{color}              
\usepackage{float}              
\usepackage{appendix}           
\usepackage{dsfont}             
\usepackage{amsmath}            
\usepackage{amssymb}            
\usepackage{gensymb}            
\usepackage[colorlinks=true,linkcolor=azulon,citecolor=violet,urlcolor=verdoso]{hyperref}
\usepackage{array}              
\usepackage{authblk}

\usepackage{microtype}

\definecolor{notas}{rgb}{1, 0.4, 0.}
\definecolor{azulon}{rgb}{0,.33,.61}
\definecolor{verdoso}{rgb}{0.1,0.4,0.1}
\definecolor{violet}{rgb}{0.8,0.2,1}

\title{
PyPSA-Spain: an extension of PyPSA-Eur to model the Spanish energy system 
}

\author[1]{Cristobal Gallego-Castillo}
\author[2,3]{Marta Victoria}
\affil[1]{\small Department of Aircraft and Space Vehicles, ETSIAE, Universidad Politécnica de Madrid, Plaza Cardenal Cisneros 3, 28040, Madrid, Spain}
\affil[2]{\small Department of Mechanical and Production Engineering and iCLIMATE Interdisciplinary Centre for Climate Change, Aarhus University, 8000, Aarhus, Denmark}
\affil[3]{\small Department of Wind and Energy Systems, Technical University of Denmark, Elektrovej, 325, 2800, Lyngby, Denmark}

\begin{document}

\maketitle

\input{00_abstract}

\input{01_intro}

\input{02_pypsa-eur}

\input{03_pypsa-spain}

\input{04_impacts}

\input{05_PNIEC}

\input{06_conclusions}

\input{07_others}

\bibliographystyle{unsrt}
\bibliography{bib_file}

\appendix

\newpage

\input{99_app_Spain_Geospatial}

\newpage
\input{99_app_q2q_solar}

\newpage
\input{99_app_load}

\newpage

\ 

\newpage
\input{99_app_PNIEC_NUTS3}

\newpage

\ 

\newpage

\end{document}

%% file: 00_abstract.tex
\begin{abstract}
\small

This work presents PyPSA-Spain, an open-source model of the Spanish energy system based on the European model PyPSA-Eur. It aims to leverage the benefits of single-country modelling over a multi-country approach. In particular, several databases provided by Spanish institutions are exploited to improve the estimation of solar photovoltaic (PV) and onshore wind generation hourly profiles, as well as the spatio-temporal description of the electricity demand. PyPSA-Spain attains hourly resolution for a entire year and represents the Spanish energy system using a configurable number of nodes, while selecting around 35-50 nodes is identified as a good compromise between spatial resolution and model simplicity. 
To accommodate cross-border interactions, a nested model approach with PyPSA-Eur was used, wherein time-dependent electricity prices from neighbouring countries were precomputed through the optimisation of the European energy system.
As a case study, the optimal electricity mix for 2030 was obtained and compared with the latest update of the Spanish National Energy and Climate Plan (NECP) from September 2024.
\end{abstract}

%% file: 01_intro.tex
\section{Introduction} \label{sec_01_intro}

One of the principal actions to be taken to address the climate emergency is the decarbonisation of the electricity sector. The transformations required to attain large shares of renewable generation in the short term present a number of emerging challenges for the power system, most of which stem from the variability of the renewable resources (\textit{e.g.} wind and solar radiation) at several spatio-temporal scales. 
Accurately modelling the impact of this variability and the existing strategies to deal with it 
is therefore a necessary condition for achieving security of supply and low carbon emissions in a cost-effective manner.

PyPSA-Eur\footnote{\url{https://github.com/PyPSA/pypsa-eur}} is a state-of-the-art open energy model of the European energy system, specifically designed to analyse alternative strategies to attain timely decarbonisation. Its spatio-temporal resolution allows proper representation of the different strategies to balance renewable fluctuations and resource competition when optimising the system, including: 
(i) the spatial balancing between generation and demand through regional integration and grid transmission; 
(ii) the role of storage technologies for temporal balancing; 
(iii) the additional demand and flexibility provided by the direct electrification (\textit{e.g.} heat pumps and electric vehicles) or indirect electrification (\textit{e.g.} synthetic fuels) of other sectors including a detailed carbon accounting; 
(iv) the identification of regions for renewable capacity installation, considering the trade-off between renewable resource and the cost of grid expansion.

Initially conceived as a tool to perform analyses of the power system for a specific time horizon, PyPSA-Eur \cite{Hoersch2018} has undergone progressive developments in a collaborative framework that have rapidly extended its scope in several relevant directions. These include: (i) the incorporation of other energy carriers, including biomass, H2, methane, and liquid hydrocarbons; (ii) the extension of the model from the electricity sector to the energy sector through sector-coupling \cite{Brown2018a}, encompassing heating, transport, industry and agriculture; (iii) the inclusion of multi-horizon simulations to enable transition path analyses. As a result, PyPSA-Eur has been fruitfully utilised in recent years to address key questions pertaining to the European energy sector and its decarbonisation. For instance, in \cite{Victoria2020}, the question of whether to implement an early and steady or a late and rapid  transition pathway of the energy system decarbonisation was addressed from the cost-effectiveness standpoint. This analysis was expanded to analyse transition paths under alternative carbon budgets in \cite{Victoria2022}.
In \cite{Pedersen2022}, the consequences of halting the gas supply from Russia to Europe, within the context of the Ukraine invasion, were examined. Specifically, the study analysed the impacts on the European energy transition under the +1.5$\degree$ and +2$\degree$ scenarios.
In \cite{Glaum2023} a meshed offshore grid in the North Sea to connect floating wind and fixed-bottom wind installations is proposed and analysed. This hypothesis provides alternative scenarios for large scale wind deployment in a context of limited onshore expansion due to public acceptance issues.
In \cite{Neumann2023} the potential impacts of a European hydrogen network are investigated, by analysing the interplay between electricity grid expansion and the implementation of a H2 pipeline system in a net-zero emissions scenario.
In \cite{Zeyen2024} several standards for defining ``green hydrogen'' according to the additionality, spatial matching and temporal matching requirements were analysed. The work reveals, among others, the importance of hourly matching between renewable generation and electrolyser operation in keeping low carbon emissions, especially in a context where the electricity mix was not yet highly decarbonised.
In \cite{Rahdan2024}, the role of distributed photovoltaics within a highly decarbonised European energy system is analysed. In particular, the impact of the coupling with other distributed technologies like heat pumps and electric vehicle batteries is revealed, and the overall cost reductions due to lower needs of transmission grid reinforcements are estimated.

It is important to note that the European (\textit{i.e.} multi-country) scope of PyPSA-Eur entails certain implications and limitations. 
Firstly, the assumption of fully coordinated action between countries is implicit in the joint optimisation of all the national energy systems in Europe. While a certain degree of coordination does take place between countries belonging to the EU,\footnote{Some examples are the European CO2 target for the Paris Agreement, legislation packages like Fit for 55 and REPowerEU, or infrastructure plans like Projects of Common Interest and Projects of Mutual Interest.} the sovereignty to define energy policy strategies resides mainly at the national level. This is the case of the National Energy and Climate Plans (NECPs), in which the EU countries stablish their own CO2 targets, policies and actions as their contribution to the EU's climate and energy objectives, with little (if any) coordination between them during their definition. 
Secondly, a multi-country model requires certain degree of homogeneity in terms of resolution, quality and format of the input data. This may limit the choice of databases to those with a European or global scope.
Finally, the optimisation of a relatively large spatial domain (with a characteristic length of approximately $\sim4\,000$ km) with high spatio-temporal resolution may require high computational resources. The resolution affects the estimation of the renewable resource, but also the number of nodes employed to represent the network of each country and the aggregation level of the electricity demand. In this regard, a single-country energy system model would allow finer spatio-temporal resolutions for the same computational resources, increasing the accuracy of modelling the interplay between the renewable resource variability, the energy demand and the electricity network. Section \ref{sub_PE_methodology} provides an example of the modelled spatial scales in PyPSA-Eur when considering Europe or Spain, and includes a discussion of the extent to which an increase in resolution is positive, when the network resolution is compared to the resolution of the weather database.

The above limitations can be overcome by using a single-country energy system model. But this raises the issue of how the interconnections with neighbour countries are represented, since the exchange of renewable generation surpluses between countries is likely to play an important role in a cost-effective design of a decarbonised energy system.

Previous experience of implementing energy models at country level based on PyPSA \cite{Brown2018}, the core Python package of PyPSA-Eur, includes Poland \cite{Instrat}, Ireland \cite{McMullin2021} and Germany \cite{Nebel2022}. These models mimic basic aspects of PyPSA-Eur, but are based on the copper-plate hypothesis, \textit{i.e.} the network consists of a single node. Thus, the powerful modelling capabilities of PyPSA-Eur based on spatial resolution are not included.
\cite{Nebel2022} analysed the impact of different national CO2 targets on the optimal mix in Germany assuming an isolated power system, which represents a limitation for a well-connected country in North Central Europe.
The project PyPSA-Ariadne \cite{PyPSA_Ariadne} implements a German energy system model based on PyPSA-Eur. The model allows a high resolution representation of the network (up to 50 nodes). Interaction with neighbouring countries is implemented by representing them with one node per exterior country. This means that coordination in energy planning between the involved countries is assumed, as their energy systems are jointly co-optimised.
PyPSA-GB is a recent model for the Great Britain's  power system \cite{Lyden2024}. It considers spatial resolution (29 nodes) and makes extensive use of databases from national institutions. Interaction with neighbouring countries is implemented assuming historical flows for existing transmission capacity, and through HVDC links with a fixed electricity price for future transmission capacity associated with planned projects. According to the authors, the use of fixed electricity prices for the interconnections is a limitation as the dynamic interaction between countries is not properly captured. An alternative approach is to use results from exogenous models with a larger spatial domain, such as interconnection power flows or variable electricity prices. This is implemented here for the development of a Spanish energy system model.
Table \ref{tab_single_models} summarises the main features of PyPSA-based single-country energy system models. All of them are compatible with a temporal resolution of one hour.

\begin{table}[h!]
\centering
\footnotesize
\begin{tabular}{lllllc}
\hline
\textbf{Country} & \textbf{Model/Project} & \textbf{Basis}  & \textbf{Spatial resolution} & \textbf{Interconnections} & \textbf{Ref.}\\ 
\hline
Poland  & PyPSA-pl  & PyPSA & 1 node & \textit{Fixed capacity}  & \cite{Instrat} \\
Ireland & OESM-IE   & PyPSA & 1 node & \textit{Store}  &  \cite{McMullin2021} \\
Germany & Cire      & PyPSA & 1 node & \textit{Isolated}  &  \cite{Nebel2022} \\
Germany & PyPSA-Ariadne & PyPSA-Eur & Up to 50 nodes$^\dagger$ & \textit{Co-optimised}   & \cite{PyPSA_Ariadne} \\
GB & PyPSA-GB  & PyPSA-Eur & 29 nodes & \textit{Fixed flow/price}  & \cite{Lyden2024} \\
Spain & PyPSA-Spain & PyPSA-Eur & Configurable$^\ddagger$ & \textit{Hourly price}  & This work \\
\hline
\end{tabular}
\caption{Single-country PyPSA-based energy system models. \textit{Fixed capacity}: fixed generation capacities are assumed in the neighbouring countries. \textit{Store}: neighbouring countries are represented as energy stores, with net balance between exports and imports on an annual basis. \textit{Isolated}: no interconnections with neighbouring countries. \textit{Co-optimised}: the energy exchanges result from the co-optimised capacities and dispatch with neighbouring countries. \textit{Fixed flow/price}: for existing interconnections, historical flows are used; for new interconnections, energy exchanges are optimised from assuming constant electricity prices in the neighbouring countries. \textit{Hourly price}: time dependent prices, see Section \ref{subsec_interconnections} for details. $^\dagger$ of which 30 for Germany. $^\ddagger$ Lower than the number of nodes considered in the model of the real network (700); this work considers up to 100 nodes.}  \label{tab_single_models}
\end{table}

\begin{figure}[!ht]  
\centering
\includegraphics[width=12cm]{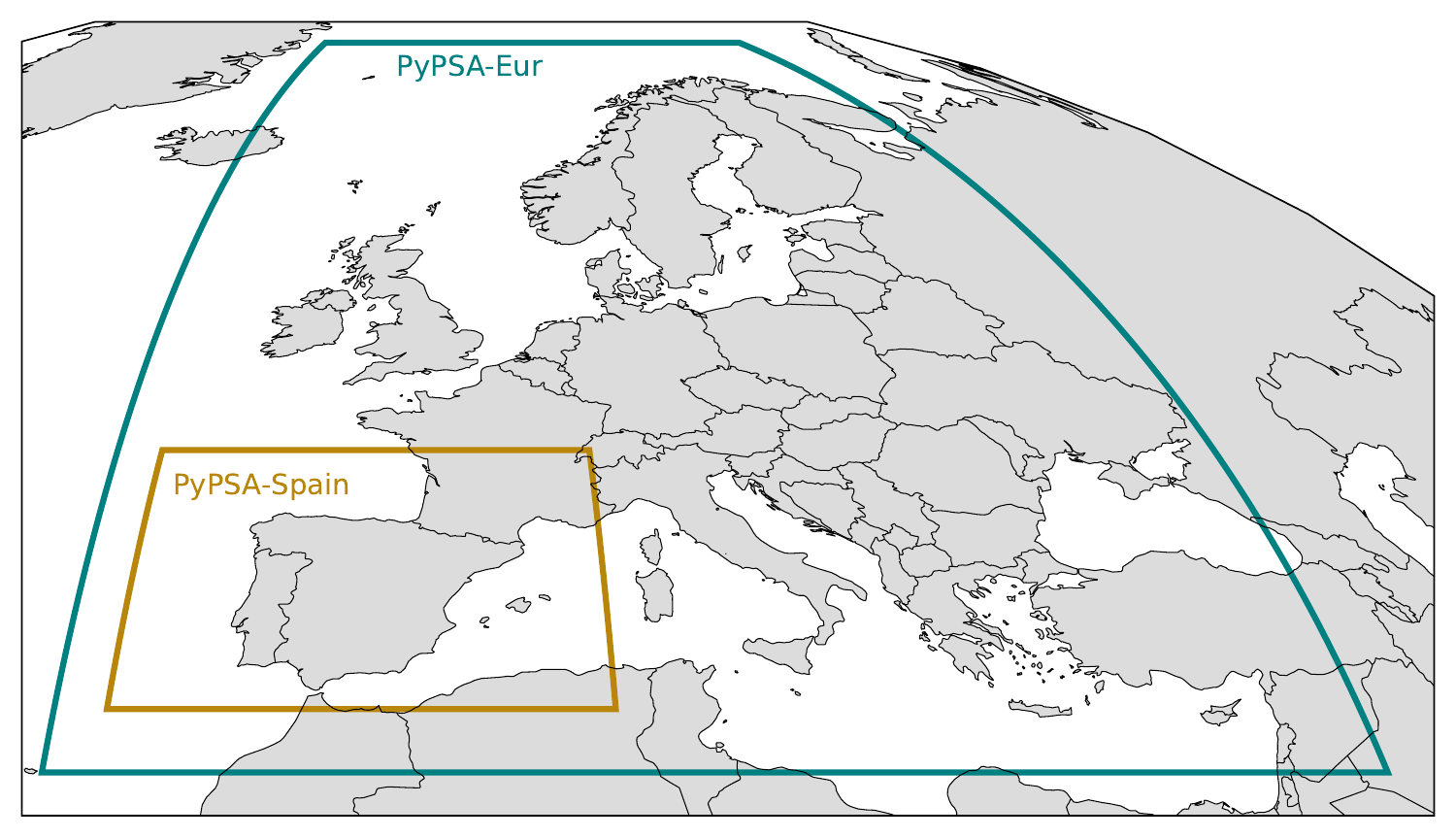}
\caption{Spatial domain considered in PyPSA-Eur and PyPSA-Spain.} \label{fig_nested_maps}
\end{figure}

The approach proposed in this paper is inspired by the strategy adopted in weather forecast. This strategy consists in using nested models: large domain and low resolution weather models (Global Circulation Models, covering the whole atmosphere) provide initial and boundary conditions to small domain and high resolution models (Limited Area Models, covering a specific spatial domain, \textit{e.g.} regional climate models such as EURO-CORDEX). This combination of models provides better weather forecasts than those obtained with each model separately. 
Similarly, we propose the implementation of a single-country energy system model (PyPSA-\textit{country}) operating in a scenario generated with a multi-country model (PyPSA-Eur). Both modelling levels work with the similar variables and methodologies, but different spatio-temporal resolution. 
As a particular case, this paper presents PyPSA-Spain, an extension of PyPSA-Eur focused on the Spanish energy system. Figure \ref{fig_nested_maps} shows the nested domains considered in PyPSA-Eur and PyPSA-Spain.

PyPSA-Spain is an up-to-date fork of PyPSA-Eur, which ensures that new developments, corrections and improvements implemented in PyPSA-Eur are easily integrated. In addition, PyPSA-Spain includes a number of original functionalities that improve the representation of the Spanish energy system. Prominent among them are: (i) improved estimation of onshore wind and solar PV generation profiles, fitted and validated with historical data; (ii) improved spatio-temporal description of the electricity demand, obtained through historical demand profiles at NUTS 2 and NUTS 3 level; (iii) a flexible modelling of interconnections with neighbour countries, with time-dependent price-based electricity exchanges. The first version of PyPSA-Spain presented here focuses on the power system. Further developments will consider improvements in other features included in PyPSA-Eur, such as sector coupling and multi-horizon modelling.

The above features make PyPSA-Spain a state-of-the-art model that improves existing approaches to modelling the energy system in this country. 
Time-resolved analyses of the Spanish power system have been carried out in the past \cite{Galbete2013, Linares2017,Victoria2019,Gallego-Castillo2021a,Bonilla2022}. At the governmental level, at least two relevant analyses have been performed: the one included in the Experts Commission on Energy Transition report commissioned by the Spanish government in 2017 \cite{CETE2018} and the Spanish National Energy and Climate Plan (NECP, or PNIEC by its Spanish acronym) \cite{MITERD2024}. All these studies are characterised by the adoption of the single-node hypothesis, which neglects the relevance of spatial resolution by assuming infinite capacity of the transmission network. As discussed above, this assumption has several limitations when modelling power systems with high share of renewable sources.

The reminder of this paper is structured as follows:
Section \ref{sec_02_pypsa-eur} contains an introductory description of PyPSA-Eur.
In Section \ref{sec_03_pypsa-spain}, a description of the new functionalities implemented in PyPSA-Spain is included.
Section \ref{sec_04_impacts} assesses the impact of the main functionalities introduced on the optimal configurations of the power system. 
In Section \ref{sec_05_PNIEC}, PyPSA-Spain is used to obtain optimal energy system configurations to achieve the 2030 decarbonisation target. The results are compared to the recently updated version of the Spanish NECP.
Finally, the main conclusions are summarised in Section \ref{06_conclusions}.

%% file: 02_pypsa-eur.tex
\section{An overview of PyPSA-Eur} \label{sec_02_pypsa-eur}
    
PyPSA-Eur is an open energy model of the European energy system. Its spatio-temporal resolution makes it suitable for modelling high shares of renewable energy. The scope of the model includes both capacity and dispatch optimisation. The model combines information at several levels to formulate an optimisation problem with constraints to obtain cost-optimal configurations of the energy system for a given energy demand scenario and decarbonisation target. The problem formulation relies on the assumptions of long-term market equilibrium, perfect competition and foresight. The results include optimal generation and storage capacities, grid expansion requirements, optimal dispatch of controllable generation and storage technologies, electricity prices in every node and required CO2 tax, among others. See Supplementary Note S15 in \cite{Victoria2022} for a mathematical description of the objective functions and constraints.

\subsection{Input data}

This section provides a summary of the different input data employed by PyPSA-Eur. A key aspect is that all the employed data come from open databases\footnote{\href{https://github.com/PyPSA/pypsa-eur/blob/master/doc/data_sources.rst}{https://github.com/PyPSA/pypsa-eur/blob/master/doc/data\_sources.rst}}.

\begin{enumerate}
    
    \item Power grid: data of high-voltage transmission network in Europe from ENTSO-E and OpenStreetMap.
    
    \item Electricity demand: historical hourly time series aggregated by country from ENTSO-E.
    
    \item Historical conventional and renewable generation capacity: data of power plants from ENTSO-E, WEPP, OPSD, etc., combined and curated by the Python package powerplantmatching, see \cite{Gotzens2019}. For countries where the location of renewable plants is not available, a heuristic spatial distribution of the capacity based on the annual capacity factor is implemented.
    
    \item Renewable resource: atmospheric data from reanalysis and satellite datasets (ERA-5 and SARAH), with $0.25\degree \times 0.25\degree$ lon-lat resolution, and temporal resolution of one hour.

    \item Land-use data: includes Natura 2000 and CORINE Land Cover database. The former defines protected areas, and the latter provides 44 land cover and use classes with a spatial resolution of 100 meters. Both datasets are employed to limit eligible areas for renewable energy deployment, see an example for Spain in Appendix \ref{app_Spain_Geospatial}.

    \item Techno-economical data: includes a set of relevant parameters (capital cost, fixed and variable Operation and Management (O\&M) costs, lifetime, efficiency, CO2 intensity, etc.) of the considered technologies (generation, storage, transmission, etc.). See details in \url{https://github.com/pypsa/technology-data}.
    
\end{enumerate}

In addition to these data sets, the model requires other configurable inputs that have to be provided by the user:

\begin{itemize}
    
    \item Setting constraints such as a global CO2 limit, either for a single year or for a transition path, and a grid expansion limit, to account for public acceptance issues.
    
    \item Parameters to simplify the power system, like number of clusters in which the transmission network is simplified, wind turbine model, solar PV panel angle, etc.
    
\end{itemize}

Further input data and configurable parameters are required for the sector-coupling layer of PyPSA-Eur. For the sake of brevity, these are not described here, since the first version of PyPSA-Spain presented in this paper covers only the power system modelling.

\subsection{Methodology} \label{sub_PE_methodology}

A detailed review of the methodology implemented in PyPSA-Eur is beyond the scope of this paper. Only a summary is provided here, with special focus on some aspects of the workflow to facilitate the understanding of the new functionalities implemented in PyPSA-Spain. For more details, the reader is referred to \cite{Hoersch2018} and \href{https://pypsa-eur.readthedocs.io}{https://pypsa-eur.readthedocs.io}.

\subsubsection*{Spatial scales}

One of the key features of PyPSA-Eur is the spatial resolution in the modelling of the transmission network. 
Two different scales can be identified, here referred to as the real network scale and the clustered network scale. 

\begin{figure}[!ht]  
\centering
\includegraphics[width=14.35cm,trim={0 70 0 0},clip]{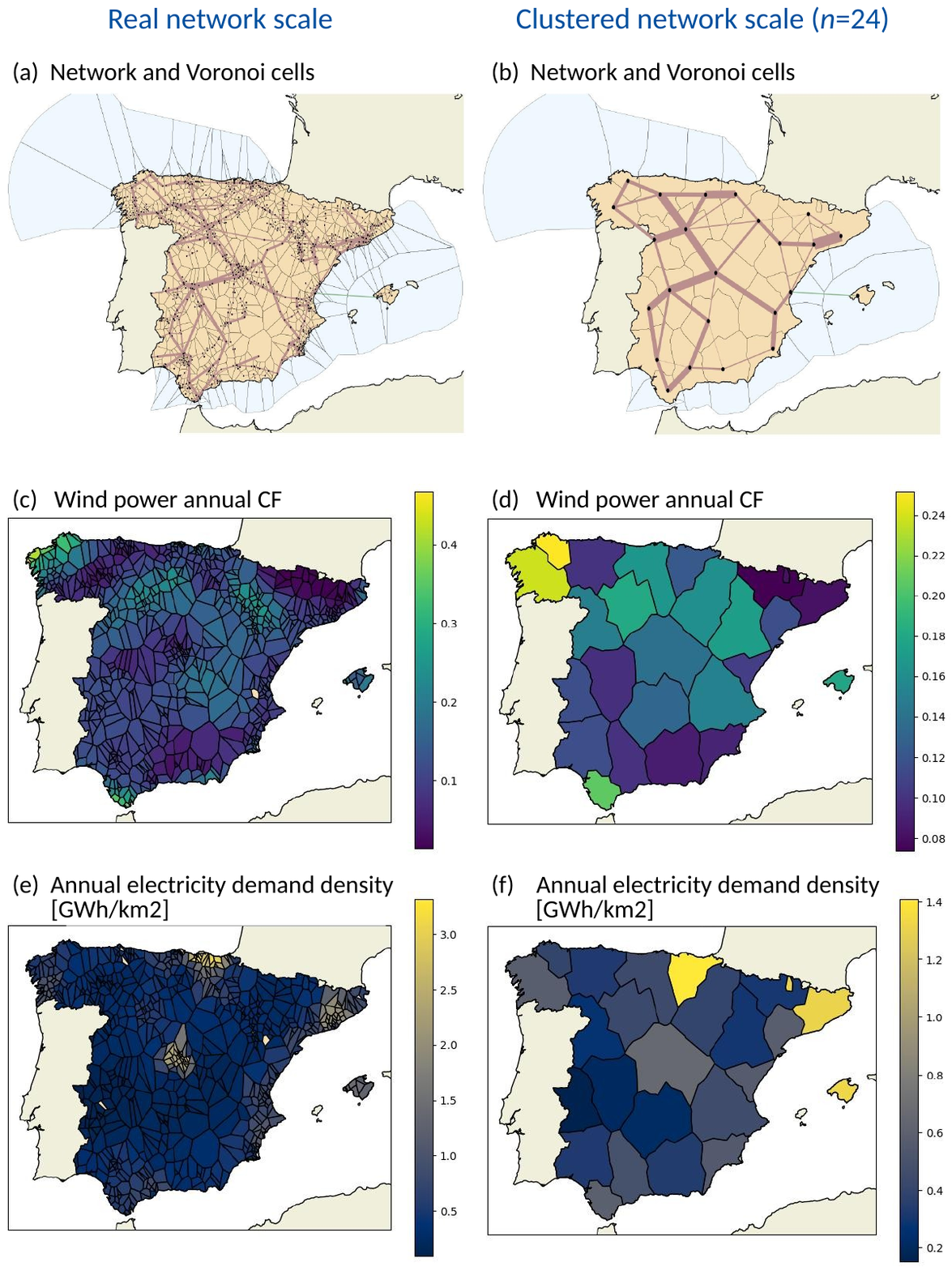}
\caption{Transmission network and Voronoi cells (top), wind power annual capacity factor (middle) and annual electricity demand density (bottom) at the real network scale (left) and for the clustered network scale for $n=24$ nodes (right), calculated with PyPSA-Eur for Spain in 2022.} \label{fig_scheme}
\end{figure}

The real network scale is a comparatively high resolution scale. It is based on the HV transmission network (input data from ENTSO-E or OpenStreetMap), which is used to define nodes and grid lines (AC and DC). The points over land closest to each node defines a geographical area known as Voronoi cell. 
This scale comprises about $700$ nodes and Voronoi cells for the case of Spain,
see Figure \ref{fig_scheme}-a.

\begin{figure}[!ht]  
\centering
\includegraphics[width=14.5cm]{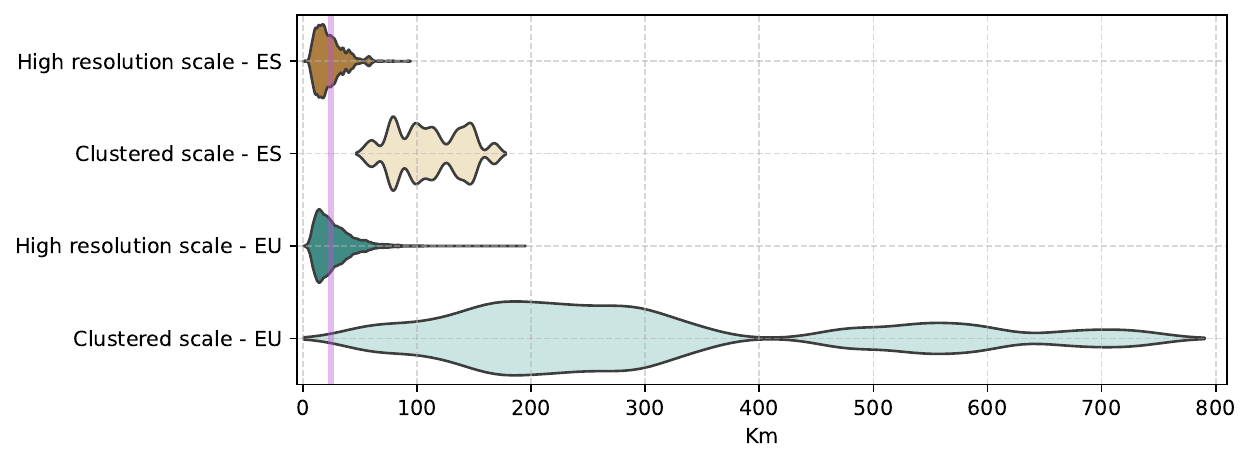}
\caption{Distribution of characteristic lengths (squared root of the Voronoi cell areas) for the real network and clustered network spatial scales in Spain and Europe. The clustering was done with 38 nodes in both cases. The purple line represents the size of the grid cell in the ERA-5 reanalysis database.} \label{fig_spatial_scales}
\end{figure}

The clustered network scale is a lower resolution scale, and it is based on the clustered network resulting from applying a $k$-means algorithm to the real network with a target number of nodes, see an example for 24 nodes in Figure \ref{fig_scheme}-b. This clustering reduces the computational burden. Figure \ref{fig_spatial_scales} compares the two models when Europe or Spain are considered as domains. Each row shows the range of characteristic lengths for each case, represented by the statistical distribution of the square root of the Voronoi cell areas. In both cases, 38 nodes were considered, since this is the minimum number of nodes required for the case of Europe (one node per country or region of European countries belonging to separate synchronous zones). It can be seen that, after clustering, the characteristic lengths  obtained in Spain are much smaller than those obtained in Europe. This can help prevent an excessive smoothing of the renewable resource, in particular for the wind generation.
Preserving smaller characteristic lengths at the clustered network scale also helps capture better the spatial distribution of the electricity demand, and potential bottlenecks in the electricity network. The vertical magenta line represents the characteristic length of a $0.25\degree \times 0.25\degree$ reanalysis grid cell (ERA5 horizontal resolution). It suggests a low limit of the spatial resolution for proper estimation of renewable generation, as smaller Voronoi cells would not lead to better estimation of wind and solar hourly capacity factors. However it is important to note that Voronoi cells of similar or smaller size than weather grid cells can be problematic in areas where the renewable resource may experience strong spatial gradients. For example, consider a weather grid cell in a coastal location where the wind speed obtained is high due to offshore conditions. This data is also assigned to inland areas within the weather grid cell. In such a situation, if the Voronoi cell is smaller than the weather grid cell, the corresponding hourly capacity factor for onshore wind power in that Voronoi cell would be overestimated, as it is calculated using only information from the mentioned weather grid cell. Therefore, increasing the resolution of the network should only be considered as a positive measure as long as it does not exceed the resolution of the weather data, especially in the clustered network, which is employed in the optimisation stage (see below).

\subsubsection*{Calculations at the real network scale}

Wind and solar hourly capacity factors are initially obtained for each node of the real transmission network. This is done by combining weather data from weather grid cells that overlap the Voronoi cell with energy system models, \textit{i.e.} wind turbine or PV panel model. For example, a wind turbine model includes the wind power curve, which relates the non-linear relationship between wind speed at hub height and the power output of the wind turbine. An important underlying hypothesis at this stage is that renewable capacity is distributed across weather grid cells contained in the Voronoi cell proportionally to their annual capacity factor, which is consistent with the idea that renewable capacity is deployed according to the resource. In addition, the estimation of the hourly capacity factor in a Voronoi cell is performed taking into account only its eligible area. The eligible area excludes Natura 2000 protected areas, and specific land covers and uses according to the CORINE Land Cover classification, see Appendix \ref{app_Spain_Geospatial} for details. More details on this process can be found in \cite{Hoersch2018}. Internally, PyPSA-Eur uses the Python package Atlite to calculate the hourly capacity factors.
As an example, Figure \ref{fig_scheme}-c shows the resulting annual capacity factor (the average of the hourly capacity factors throughout the year) at each node for onshore wind in Spain. In addition, the historical installed capacity of each technology is set at each node. For wind and solar PV technologies, if the exact location of the plants is not available (as in Spain), a heuristic distribution across the nodes is made proportional to its annual capacity factor. The underlying assumption is that renewable generation capacity is deployed more intensively at locations with a higher resource. The generation time series of wind and solar at each node can be obtained by multiplying the hourly capacity factor by the installed capacity. Finally, a maximum capacity for solar PV and wind power is computed for each Voronoi cell. This cup is used during the optimisation phase to limit the expansion of renewable capacity. To calculate this limit, a power density in terms of MW/km$^2$ is defined in the configuration file for each renewable technology.

In addition, an electricity demand time series is attached to each node. It is obtained by re-scaling the country electricity demand time series. Thus, all nodes within a country have the same hourly profile. The annual electricity demand of every node is proportional to the population and the gross domestic product (GDP), obtained from data at NUTS 3 level, according to predefined coefficients obtained from a regression analysis of the per country data in Europe. These coefficients are 
$$gdp = 0.60 ~~,~~ pop=0.40,$$ 
\noindent see \cite{Hoersch2018} for details.
Figure \ref{fig_scheme}-e shows the resulting annual electricity demand density of each node in Spain.

\subsubsection*{Calculations at the clustered network scale}

A clustered version of the network is obtained by aggregating the nodes using $k$-means clustering. Figure \ref{fig_scheme}-b shows the case for $n=24$ nodes. The information from the high resolution scale (renewable profiles, installed capacity, electricity demand time series, etc.) is aggregated and attached to the nodes of the clustered network. Figures \ref{fig_scheme}-d and \ref{fig_scheme}-f show the corresponding versions of figures \ref{fig_scheme}-c and \ref{fig_scheme}-e for the clustered network. The impact of the aggregation can be seen on the maximum annual capacity factor, which decreases from 0.45 at the high-resolution model to 0.25 at the clustered model.

\subsubsection*{Joint capacity and dispatch optimisation}

PyPSA-Eur provides cost-effective configurations of the energy system by means of an optimisation problem with constraints, where the annualised total system cost computed for the clustered model is minimised. The annualised cost includes investments in generation, storage and grid expansion, as well as operating costs. The constraints are of different nature: (i) technical constraints, to ensure security of supply and compliance with Kirchhof's rules of the linearised power flow at each time step; (ii) policy constraints, to keep CO2 emissions and grid expansion below given thresholds; (iii) regulatory constraints, to prevent new renewable capacity from being installed in non-eligible areas. See Supplementary Nota S15 in \cite{Victoria2022} for a mathematical description of the objective function and constraints.

%% file: 03_pypsa-spain.tex
\section{Novel functionalities developed in PyPSA-Spain} \label{sec_03_pypsa-spain}

This section describes the new functionalities implemented in PyPSA-Spain. They are divided into two groups: main functionalities with relevant modifications (subsections \ref{subsec_Q2Q}-\ref{subsec_interconnections}), and additional functionalities with minor modifications (gathered in subsection \ref{subsec_others}).

\input{03_s1_q2q}

\input{03_s2_load}

\input{03_s3_interconnections}

\subsection{Additional functionalities} \label{subsec_others}
\input{03_s4_updated_esios}

\input{03_s5_updated_gdp_pop}

\input{03_s6_costs}

%% file: 03_s1_q2q.tex
\subsection{Improving the estimation of renewable hourly capacity factors} \label{subsec_Q2Q}

\begin{figure}[!ht]  
\centering
\includegraphics[width=14.5cm]{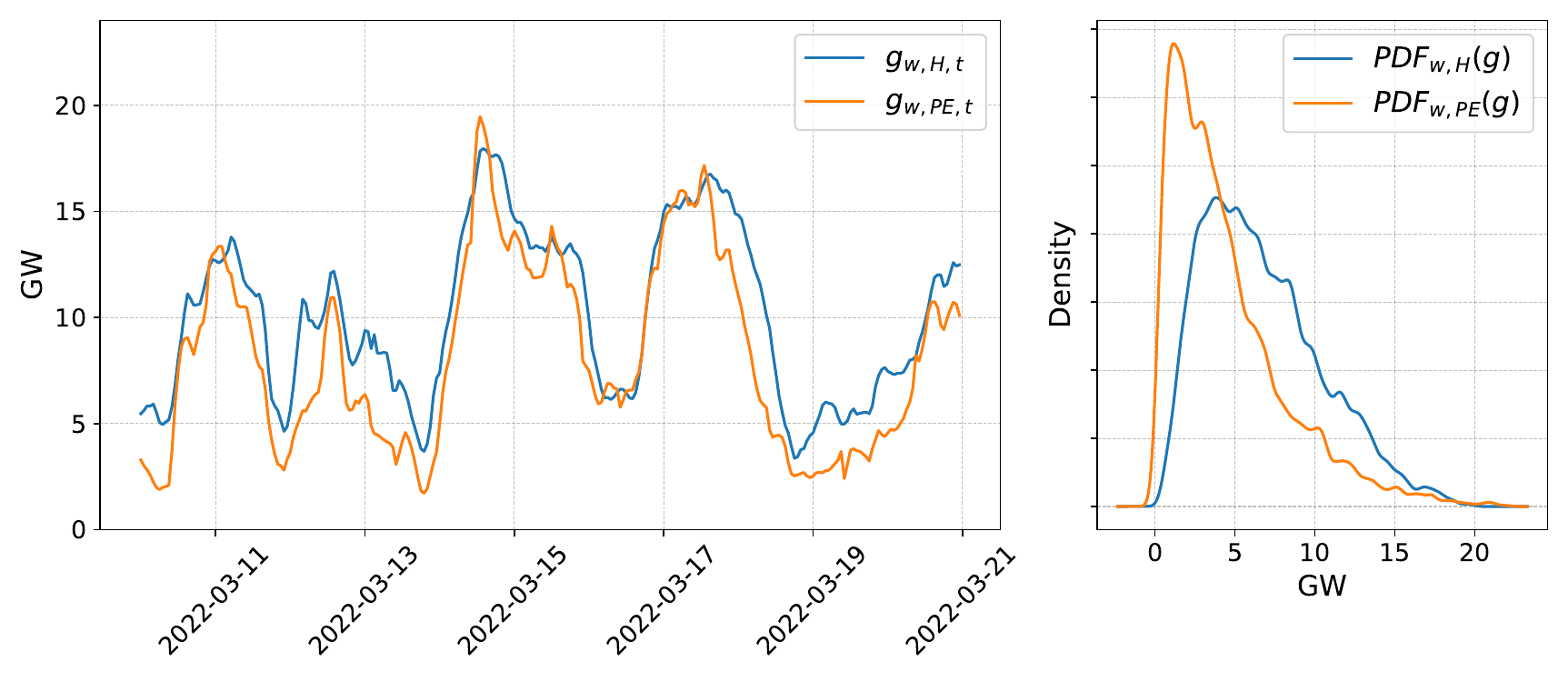}
\caption{Wind power generation in Spain in 2022. Left: time series; right: PDF. Subindices: $H$ stands for historical data, $PE$ means modelled with PyPSA-Eur.} \label{fig_onwind_ts_PDF}
\end{figure}

The wind and solar PV generation time series in PyPSA-Eur are based on the estimated hourly capacity factors at the Voronoi cells in the high-resolution model, together with the installed capacity assumed at each node. The model includes several underlying simplifications, which may lead to under/overestimate the renewable generation time series as compared with the real ones. Lets consider the case of wind power generation (a similar reasoning applies for solar generation). Figure \ref{fig_onwind_ts_PDF} shows, on the left, the historical wind power generation time series in Spain in 2022, $g_{w,H,t}$,\footnote{Data available at \url{https://api.esios.ree.es/}} and the time series estimated with PyPSA-Eur, $g_{w,PE,t}$. To compute the latter, the historical wind power capacity installed in each NUTS 2 region in Spain in 2022 was considered, see Section \ref{subsec_updated_esios} for details. On Fig. \ref{fig_onwind_ts_PDF} right, the probability density function (PDF) of both variables, denoted $PDF_{w,H}(g)$ and $PDF_{w,PE}(g)$ respectively, is shown. 
The time series plot shows that the approach implemented in PyPSA-Eur to simulate the wind power generation captures relatively well the fluctuations in time. However, the PDFs reveal a systematic underestimation of the wind power generation as compared with historical data. 

There are several potential reasons for the mismatch between the historical and modelled wind power time series, and they are related to different stages of the wind-to-power conversion chain. Some of them are:

\begin{itemize}
    \item The wind speed form reanalysis data may be subject to statistical bias in certain regions.
    \item Reanalysis data are provided at specific grid points given by the weather model, which usually are different from the actual wind farm locations. Thus, real wind farms may experience different local wind speeds than those reflected by the nearest reanalysis grid point, as wind farms are usually sited to take advantage of local effects that increase wind speed (mountain ridges, valleys, etc.).
    \item PyPSA-Eur considers one type of wind turbine, whereas a wide range of wind turbine sizes usually co-exists in a region or country.
    \item Wind power generation at a wind farm is affected by local phenomena such as wakes, which provoke a wind speed deficit at each wind turbine depending on the wind farm layout and the wind speed and direction. The wind-to-power conversion in PyPSA-Eur is done using purely the wind turbine power curve.
\end{itemize}

PyPSA-Eur includes some strategies to address wind power overestimation, such as applying a constant correction factor to the resulting hourly capacity factor, or applying a Gaussian filter to the wind turbine power curve. However, no corrections are currently implemented to deal with underestimations of wind generation.

A new methodology to improve the estimation of renewable generation profiles is implemented in PyPSA-Spain. It is based on a post-processing of the hourly capacity factors obtained for the high-resolution model. It consists in applying a quantile-to-quantile (Q2Q) transformation. In general, this transformation modifies the PDF of a random variable to match a target PDF. Ideally, the Q2Q applied to the hourly capacity factor associated to a Voronoi cell should have as a target the PDF of the historical renewable generation in the same Voronoi cell. As these data are not available, the Q2Q transform is firstly obtained at the country level using the PDF of a normalised version of the historical renewable generation data as the target. Then, the Q2Q transform is applied at the hourly capacity factor of every Voronoi cell.

Let us define the normalised version of the historical wind power generation for a specific country as 

\begin{equation}
    \Bar{g}_{w,H,t} = \frac{g_{w,H,t}}{G^{0}_{w,H}},
\end{equation}

\noindent and the normalised version of the wind power time series estimated by PyPSA-Eur for the same country as

\begin{equation}
    \Bar{g}_{w,PE,t} = \frac{g_{w,PE,t}}{G^{0}_{w,PE}},
\end{equation}

\noindent where $G^{0}_{w,H}$ and $G^{0}_{w,PE}$ are two scaling factors discussed below. The corresponding Cumulative Distribution Functions of $\Bar{g}_{w,H,t}$ and $\Bar{g}_{w,PE,t}$ are denoted by $CDF_{w,H}(\Bar{g})$ and $CDF_{w,PE}(\Bar{g})$, respectively. 

The Q2Q transform is a nonlinear function between the input normalised generation, $\Bar{g}$ and the transformed normalised generation, $g^*$, given by:

\begin{equation} \label{eq_q2q}
    \Bar{g}^* = \mathcal{Q}(\Bar{g}) =  CDF_{w,H}^{-1}(CDF_{w,PE}(\Bar{g})).
\end{equation}

The idea behind $\mathcal{Q}(g)$ is to modify the input random variable so that all of its quantiles match those of the target random variable. By considering the wind power generation computed at national level by PyPSA-Eur as the input and the historical wind power generation as the target, all potential sources of mismatch between the two variables throughout the wind-to-power chain are jointly corrected. Thus, the PDF of the transformed time series, $\Bar{g}_{w,PE,t}^*$, replicates that of the normalised version of the historical wind power generation time series, $\Bar{g}_{w,R,t}$. It should be noted that the correction is statistical in the sense that it does not require or provide any knowledge of the source of the mismatch.  

Finally, the Q2Q transformation obtained at the country level is applied to the hourly capacity factors estimated at the Voronoi cell level. This is a delicate step as the country and Voronoi scales may be quite different, and wind power generation may show different dynamics at each scale. For example, it is very unlikely that the wind power generation at the country level reaches the installed capacity, while this is not so rare for wind power generation at the Voronoi cell level, whose spatial extent can be several orders of magnitude lower. To account for this scale mismatch, an analysis on three different normalisation schemes is performed:

\begin{itemize}
    
    \item Normalisation scheme 1: based on the installed wind power capacity at country level in the considered year, $P_w$:     
    $$G^{0}_{w,R} = G^{0}_{w,PE} = P_w.$$
    
    \item Normalisation scheme 2: based on the maximum value attained by the historical and the simulated generation time series:
    $$G^{0}_{w,H} = G^{0}_{w,PE} = \max{(\max{(g_{w,R,t})},\max{(g_{w,PE,t})})}.$$

    \item Normalisation scheme 3: each generation time series is normalised with its maximum:
    $$G^{0}_{w,H} = \max{(g_{w,H,t})}$$
    $$G^{0}_{w,PE} = \max{(g_{w,PE,t})}.$$
    
\end{itemize}

\begin{figure}[!ht]  
\centering
\includegraphics[width=14cm]{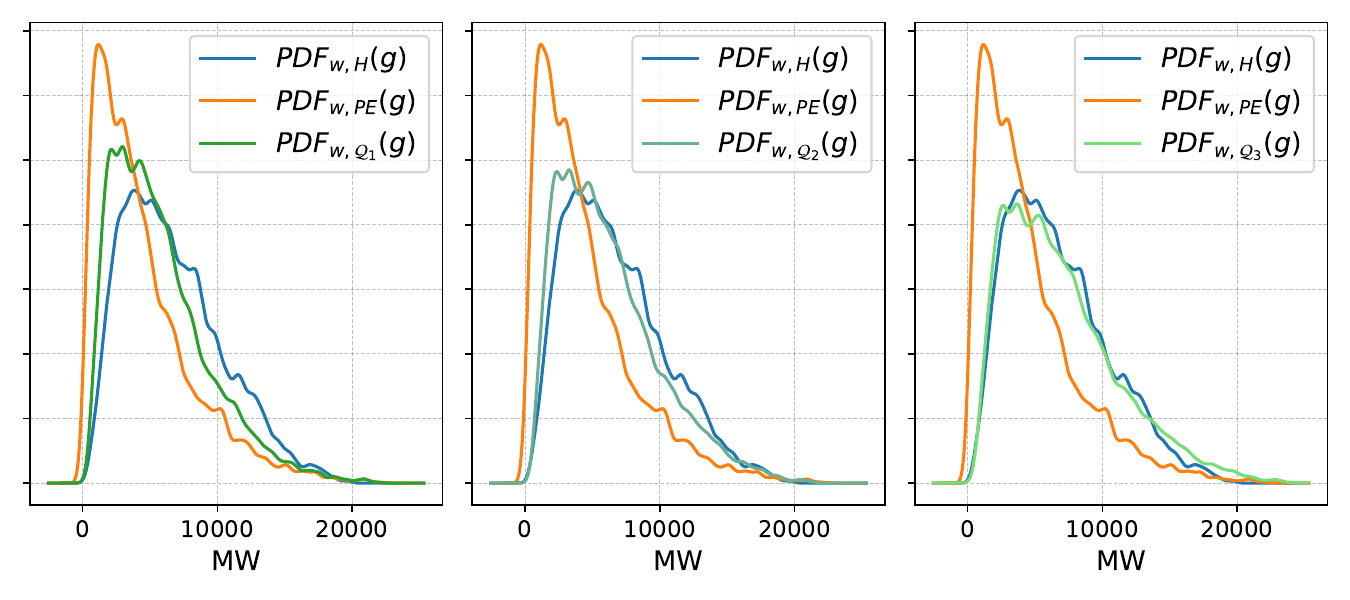}
\caption{PDFs of the wind power generation time series in Spain in 2022. Subindices: $H$ stands for historical data, $PE$ means modelled with PyPSA-Eur, and $\mathcal{Q}_i$ for $i=1,2,3$ refers to the Q2Q transform under normalisation scheme $i$.} \label{fig_q2q_onwind_pdf}
\end{figure}

The three Q2Q transforms obtained under the defined normalisation schemes, $\mathcal{Q}_1(\Bar{g})$, $\mathcal{Q}_2(\Bar{g})$ and $\mathcal{Q}_3(\Bar{g})$, are applied to the hourly capacity factors at each Voronoi cell. The resulting wind power generation time series aggregated at country level are then compared with real data to check the accuracy of the Q2Q transformation with each normalisation scheme. Figure \ref{fig_q2q_onwind_pdf} shows the PDF of the real wind power time series for Spain in 2022, $PDF_{w,H}(\Bar{g})$, the one modelled with PyPSA-Eur, $PDF_{w,PE}(g)$, and those resulting from the application of the Q2Q transform under different normalisation schemes, $PDF_{w,Q_i}(g)$ with $i=1,2,3$. It can be seen that the normalisation scheme 3 results in the best fit of the PDFs. Of particular importance is the good matching of the PDF at low levels of wind generation, as these situations are critical for sizing the storage and transmission requirements of the optimal system configuration.

\begin{figure}[!ht]  
\centering
\includegraphics[width=14cm]{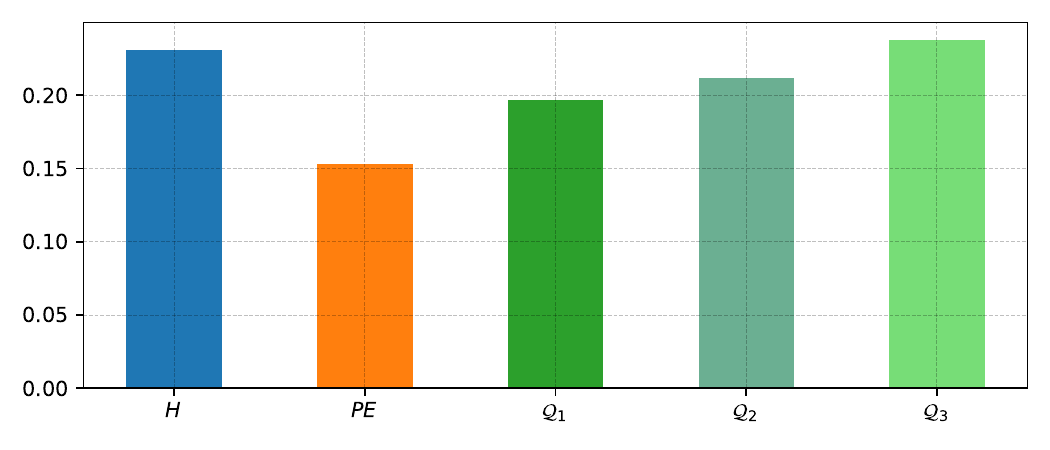}
\caption{CF of the wind power generation time series in Spain in 2022. Subindices: $H$ stands for historical data, $PE$ means modelled with PyPSA-Eur, and $\mathcal{Q}_i$ for $i=1,2,3$ refers to the Q2Q transform under normalisation scheme $i$.} \label{fig_q2q_onwind_CF}
\end{figure}

Figure \ref{fig_q2q_onwind_CF} shows the annual capacity factor (CF) obtained for each case. The Q2Q transformation under any of the proposed normalisation schemes reduces the subestimation of the CF observed in the uncorrected PyPSA-Eur values.

\begin{figure}[!ht]  
\centering
\includegraphics[width=14cm]{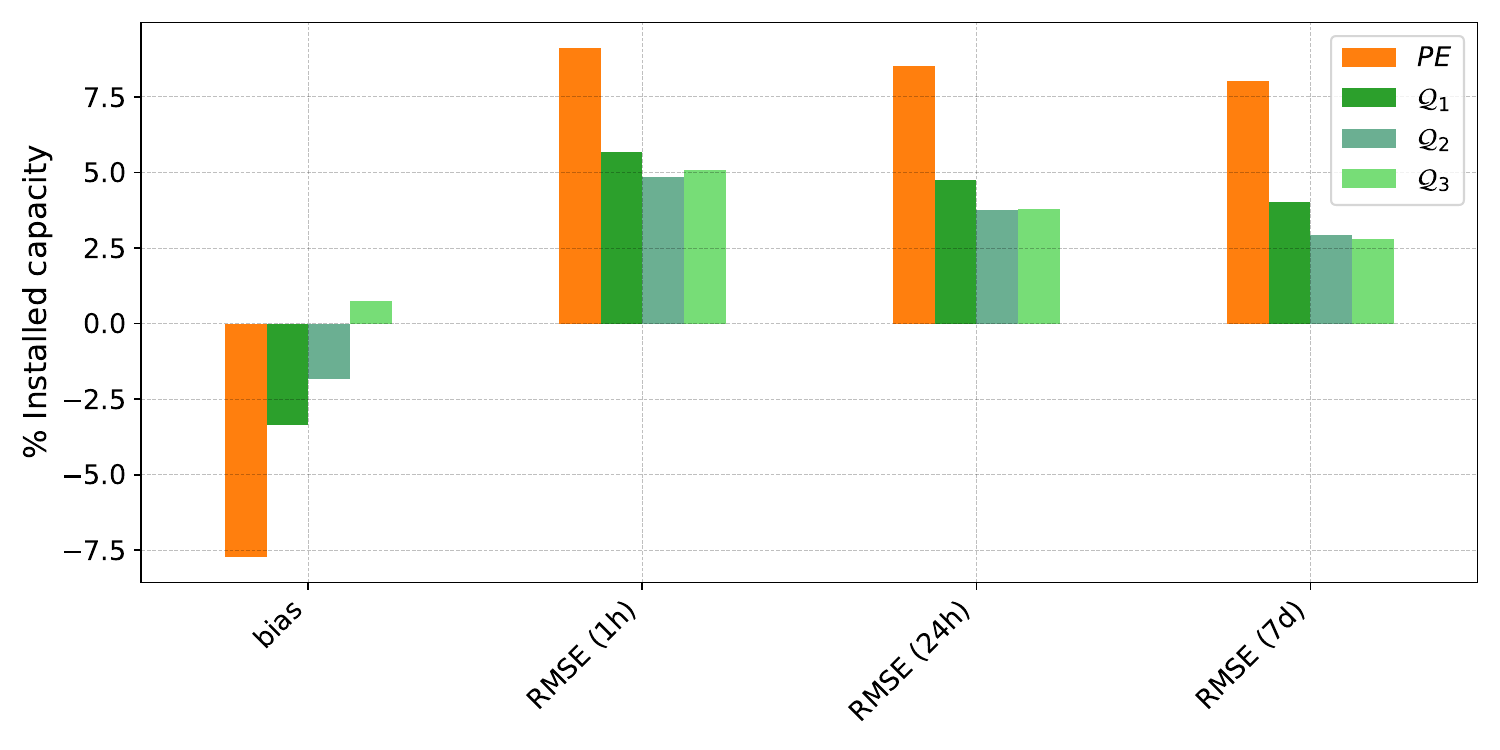}
\caption{Bias and RMSE of the simulated wind power generation time series in Spain in 2022. Subindices: $PE$ means modelled with PyPSA-Eur, and $\mathcal{Q}_i$ for $i=1,2,3$ refers to the Q2Q transform under normalisation scheme $i$.} \label{fig_q2q_onwind_RMSE}
\end{figure}

The accuracy in reproducing the historical wind power generation over time can be measured by the bias and the root mean square error. Figure \ref{fig_q2q_onwind_RMSE} shows these statistics for the different simulated cases. The RMSE is also estimated after averaging the time series with one-day and one-week time windows to capture the error across relevant wind time scales, corresponding to the diurnal cycle and the synoptic scale, respectively. It can be observed that normalisation schemes 2 and 3 provide the best performance in terms of RMSE, the latter resulting in a notably smaller bias.

Appendix \ref{app_q2q_solar} replicates this analysis for the case of the solar PV generation in Spain in 2022, concluding that the best normalisation scheme is the one considered in $\mathcal{Q}_1$. Note that the most appropriate normalisation scheme was found to be different for wind power and solar PV generation. This could be due to the fact that the above mentioned mismatch between the renewable generation at the country scale (at which the Q2Q transformation is obtained) and at the Voronoi cell scale (where the transformation is applied) operates differently depending on the renewable resource considered. This is reasonable since the spatial correlation of wind speed is quite different than that of solar radiation (in the order of hundreds of kilometres for the former and thousands of kilometres for the latter). This effect could be amplified by the well-known fact that the characterisation of sky clarity in the ERA5 reanalysis is biased towards `isotropic' skies. \textit{i.e.} there are more uniform clouds or clear skies in large regions of the reanalysis dataset than in reality.

Q2Q transformations for onshore wind and solar PV generation in Spain in 2022 under the three introduced normalisation schemes are available in the PyPSA-Spain repository. However, it should be noted that using different configuration parameters than those used to obtain the Q2Q transformations (\textit{e.g.} using a different wind turbine model, meteorological year, etc.) may reduce the performance of the Q2Q transformation. In this case it is recommended to obtain a new Q2Q transformation for the specific case following the steps here described. The details of the configuration parameters utilised to obtain the Q2Q transformations are also available in the repository.

%% file: 03_s2_load.tex
\subsection{Improving the spatio-temporal characterisation  of the electricity demand} \label{subsec_load_dist}

As described in Section \ref{sub_PE_methodology}, in PyPSA-Eur the annual electricity demand attached to each node within a country is proportional to the GDP and population of the associated NUTS 3 regions.
The hourly profile is the same for all nodes and corresponds to the hourly electricity demand time series at the national level. The importance of the hourly profile lies, among others, in the fact that it shows the relative weight of the daily peak demand (which could potentially be met by solar PV) and the seasonal patterns (which may or may not correlate with the seasonality of the wind resource).

In Spain, hourly electricity demand time series with high spatial resolution are provided by Datadis\footnote{\url{https://datadis.es/}}, a platform created by the Spanish electricity distribution companies. PyPSA-Spain takes advantage on this source to improve the spatio-temporal distribution of electricity demand.

To understand some of the temporal characteristics of electricity demand in Spain, Figures \ref{fig_load_andalucia}, \ref{fig_load_baleares} and \ref{fig_load_navarra} in Annex \ref{app_load} show the hourly electricity demand time series in 2022 in three Spanish NUTS 2 regions (Andalusia, Balearic Islands and Navarra, respectively), broken down by three economic sectors: residential, services and industrial demand. The plots are organised in a daily window, so that the line colour indicates the day of the year.

\begin{figure}[!ht]  
\centering
\includegraphics[width=5.3cm]{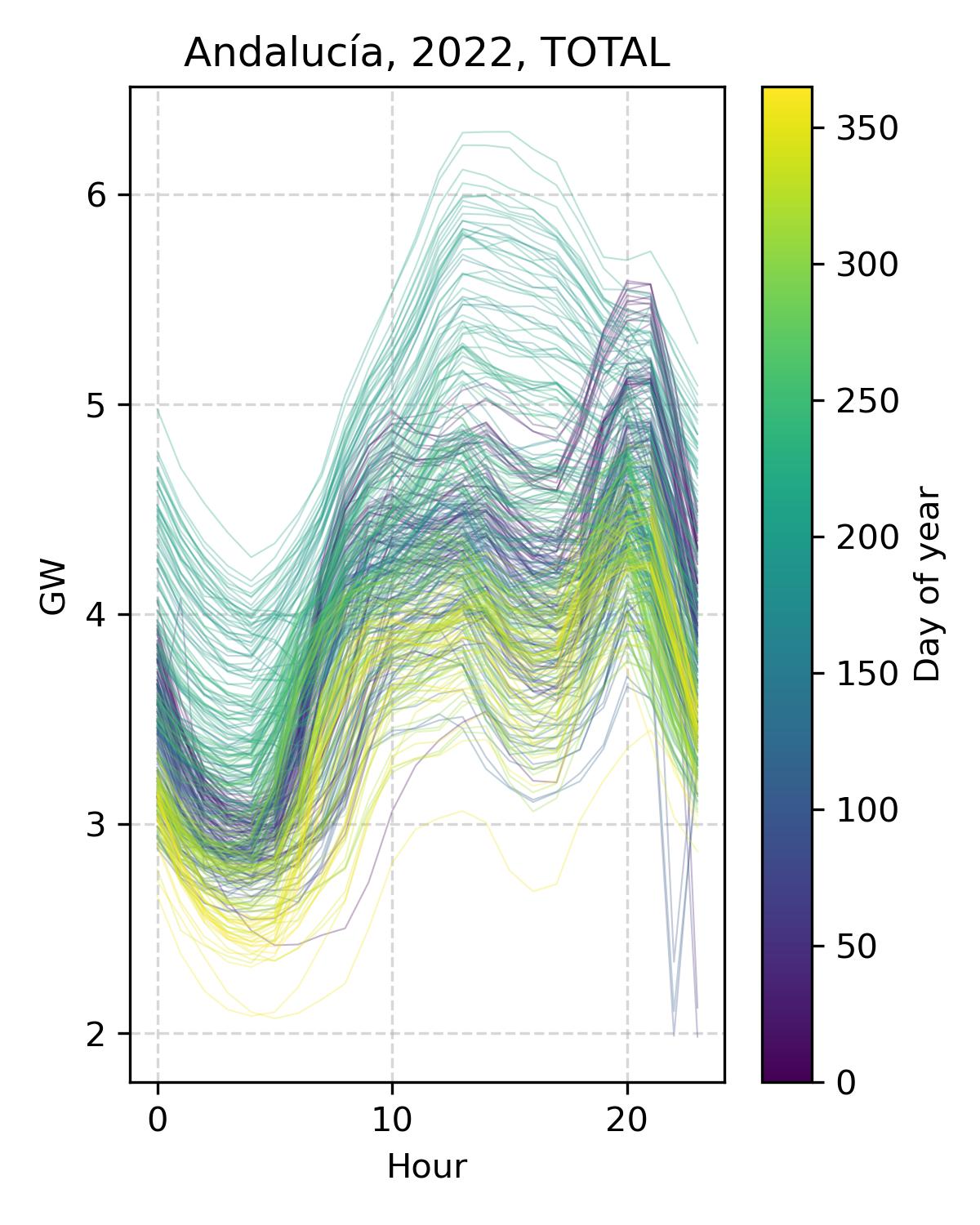}
\includegraphics[width=5.3cm]{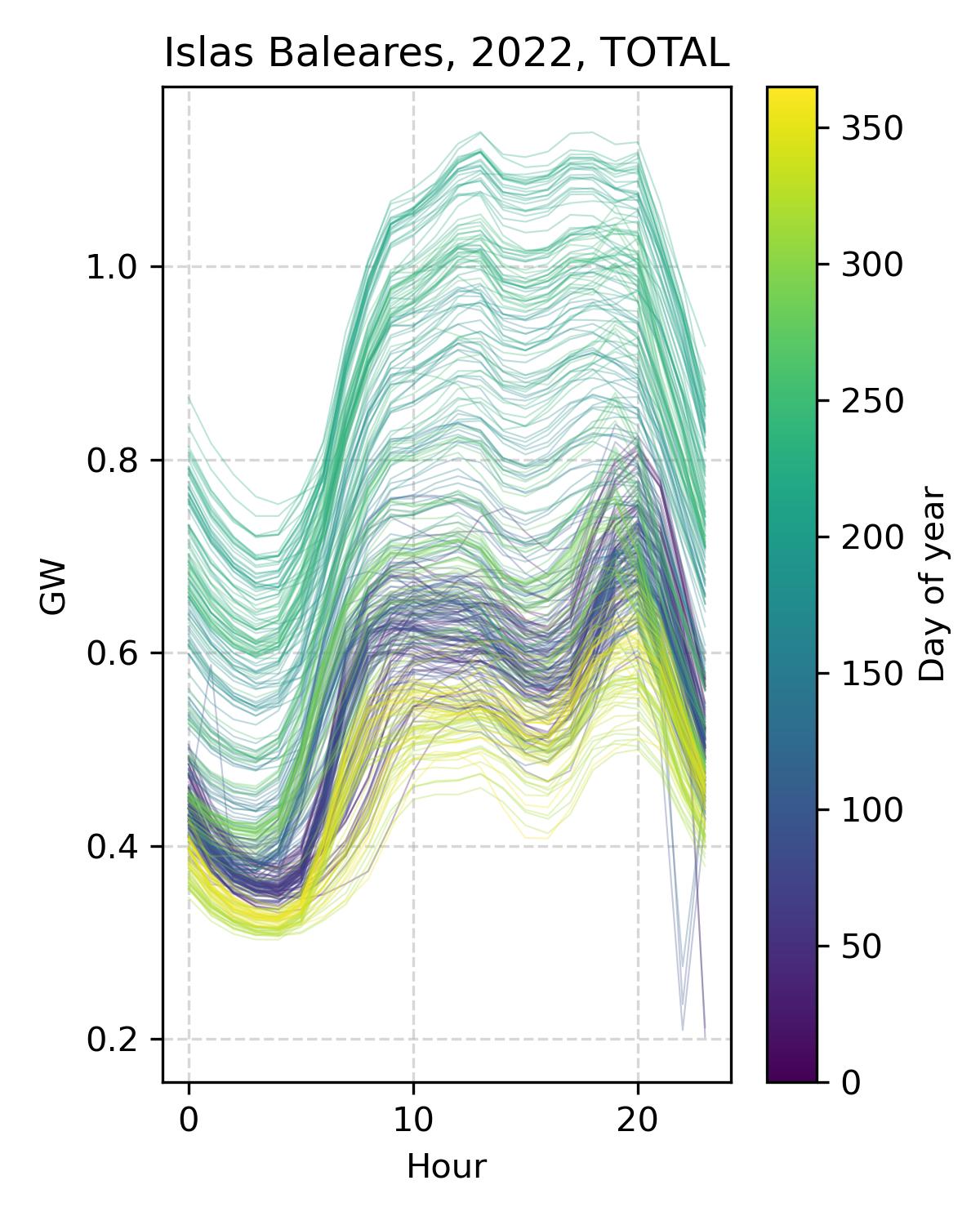}
\includegraphics[width=5.3cm]{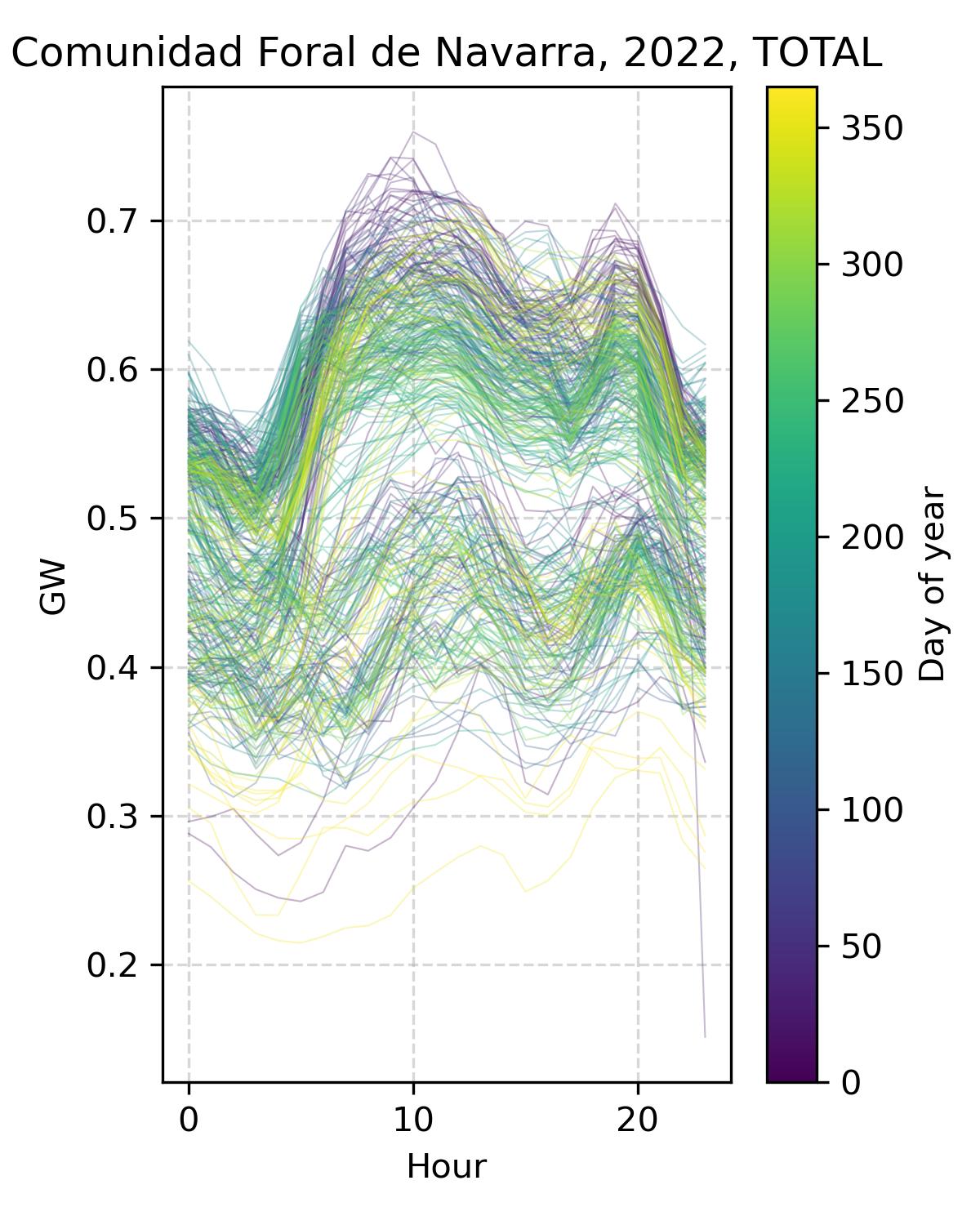}
\caption{Hourly electricity demand in three NUTS 2 regions in Spain in 2022} \label{fig_load_three_regions}
\end{figure}

It can be seen that in Andalusia, a southern region, electricity demand is roughly the same for all sectors, with high peaks in the residential sector at midday in summer and at sunset in winter. In the Balearic Islands, a highly-touristic region in the east, the main drivers are the residential and services sectors, the latter showing strong seasonality between summer and winter. Finally, in Navarre, a northern region, the electricity demand is strongly driven by the industrial sector, which has a relatively flat daily profile compared to the residential and services sectors, and a two-regime level over the year, probably due to weekdays and weekends. The corresponding plots of the total electricity demand for the three regions (\textit{i.e.} when the time series of the three economic sectors are added together for each region) are shown in Figure \ref{fig_load_three_regions}, which clearly shows the differences between the regions and over the year. This analysis shows that different regions in Spain have different temporal patterns of electricity demand, and suggests a path for improvement compared to using the same hourly profile at each node.

\begin{figure}[!ht]  
\centering
\includegraphics[width=7.5cm]{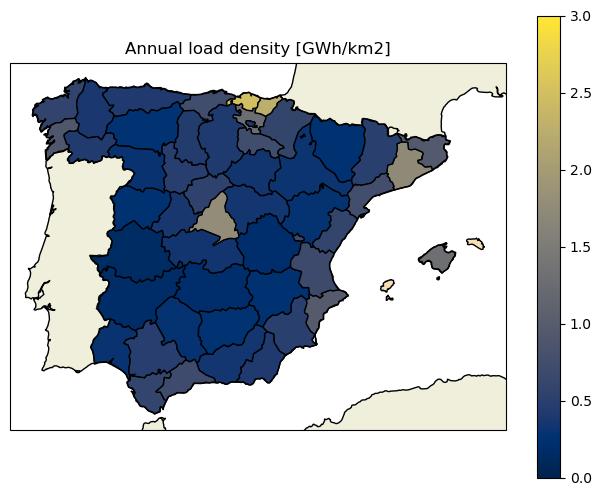}
\includegraphics[width=7.5cm]{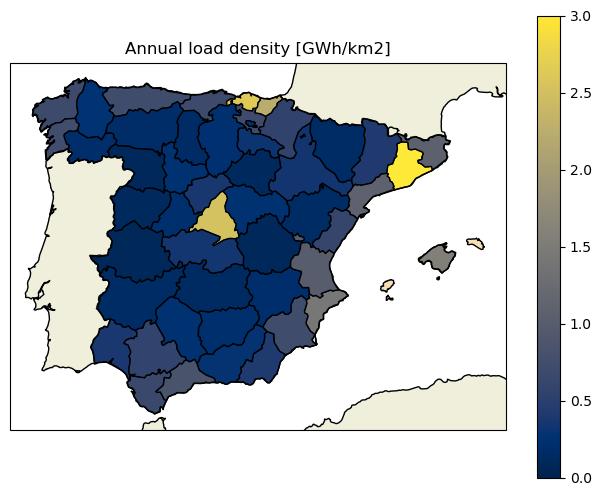}
\caption{Electricity demand density aggregated at NUTS 3 level obtained with the current methodology implemented in PyPSA-Eur (left) and PyPSA-Spain (right). The latter matches the historical data (2022).} \label{fig_annual_demand}
\end{figure}

Regarding the spatial distribution of the annual electricity demand, the methodology implemented in PyPSA-Eur (the demand at country level is shared across all the nodes) could be improved by considering the electricity demand and the nodes at NUTS 2 or NUTS 3 regions instead. This possibility is implemented in PyPSA-Spain. Figure \ref{fig_annual_demand} shows the resulting spatial distribution aggregated at NUTS 3 level obtained with PyPSA-Eur (left) and PyPSA-Spain (right). While the latter coincides with the employed historical data, the former shows a more uniform distribution of electricity demand density, underestimating in particular the high annual electricity demand density in the central (Madrid) and north-eastern (Catalonia) regions. Note that the same colour scale is used in both figures to facilitate comparison.

In PyPSA-Spain, the electricity demand is provided as input in terms of the following variables:

\begin{itemize}
    \item $\Bar{d}_{i,s,t}$, the normalised hourly demand time series of the economic activity $s$ in region $i$. Index $t$ represents the hour of the year. The mathematical definition of $\Bar{d}_{i,s,t}$ is 
    $$ \Bar{d}_{i,s,t} = \frac{d_{i,s,t}}{<d_{i,s,t}>},$$ where $d_{i,s,t}$ is the hourly demand time series for sector $s$ and region $i$, and $<d_{i,s,t}>$ is its annual average.
    \item $w_{i,s}$, the percentage of the annual electricity demand corresponding to economic sector $s$ in the region $i$.
    \item $D$, the total annual demand taking into account all the economic sectors and regions.
\end{itemize} 

From this, the hourly demand time series in region $i$, $d_{i,t}$, is obtained as follows:

$$ d_{i,t} = \frac{D}{8760} \cdot \sum_s w_{i,s} \cdot \Bar{d}_{i,s,t}.$$

Given that, the electricity demand time series $d_{i,t}$ is shared proportionally across the nodes belonging to region $i$.

The variables $\Bar{d}_{i,s,t}$ and $w_{i,s}$ corresponding to 2022 are available in the PyPSA-Spain repository for NUTS 2 and NUTS 3 regions. The variable $D$ is set in the configuration file. It is noted that the customisation of the three variables allows for creating new electricity demand scenarios with a high degree of flexibility.

%% file: 03_s3_interconnections.tex
\subsection{Modelling interconnections with neighbouring countries} \label{subsec_interconnections}

A power system with a high share of renewable generation benefits from interconnections between countries to manage excess/deficit of renewable generation. However, country energy system models are sometimes built on the basis of isolated systems. This is the case in PyPSA-Eur when only one country is selected.

The functionality described in this section models the interconnections of the Spanish power system with France and Portugal.
Each interconnection with a neighbouring country is modelled through the addition of several elements, namely:

\begin{itemize}
    \item A node located in the border (case of inland interconnection) or in the shore (submarine interconnection).
    \item A link between the aforementioned border node and the closest node of the clustered grid.
    \item An electricity demand attached to the border node. This demand is a virtual representation of the total demand in the neighbouring country, and provides a means of accounting for electricity exports from the Spanish power system.
    \item A generator attached to the border node, with zero capital cost, and a variable generation cost equal to the electricity price in the neighbouring country (different for each hour of the year). This generator is a virtual representation of the electricity market in the neighbouring country, and provides a means of accounting for electricity imports into the Spanish power system.
\end{itemize}

The electricity price time series for Portugal and France serve as inputs to the model and are used during the optimisation process to determine the direction of electricity flow in the interconnections at each time step, based on price arbitrage. In the hours when the electricity price at the Spanish node is cheaper than at the border node, the electricity is exported to cover the virtual demand in the border node, and the generation in the border node is reduced. In the opposite case, electricity generated with the virtual generator is imported to the Spanish node. The generation cost time series employed for France and Portugal have been previously obtained by optimising the European energy system using PyPSA-Eur under a specific carbon target consistent with the one considered in PyPSA-Spain. Consequently, this approach provides the mechanism by which the low resolution multi-country model (PyPSA-Eur) provides the appropriate context for the optimisation of the high resolution single-country model (PyPSA-Spain).

\begin{figure}[!ht]  
\centering
\includegraphics[width=12cm]{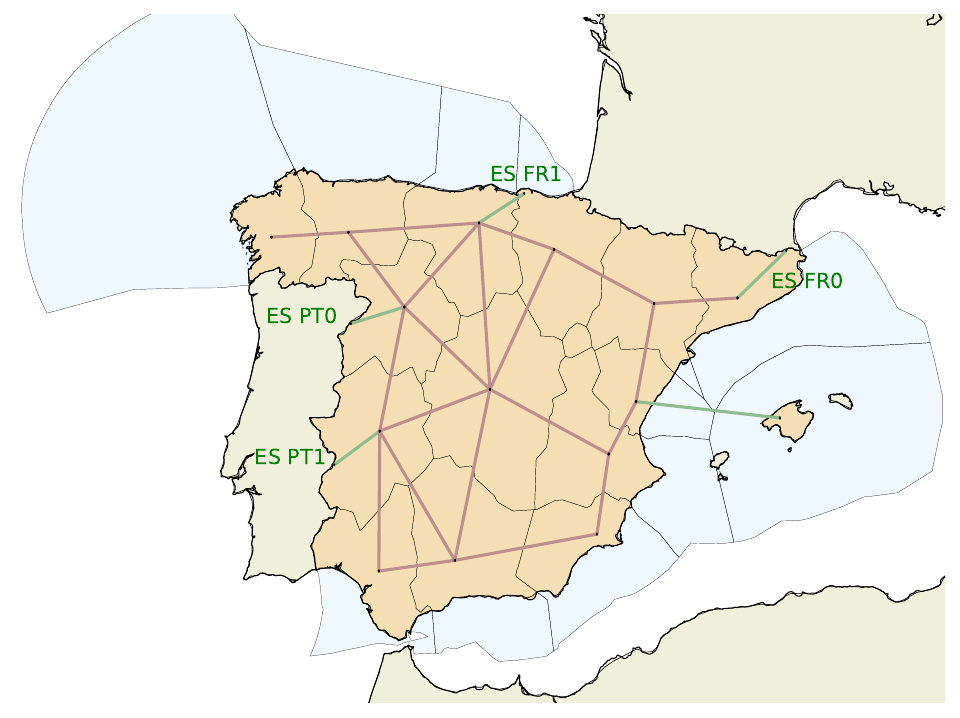}
\caption{Clustered network with 15 nodes and the default interconnections included in PyPSA-Spain.} \label{fig_interconnections}
\end{figure}

The number and characteristics of the interconnections can be customised in the configuration file by the user. The default values are described in Table \ref{tab_interconnections}. While the interconnections with France correspond to real projects, interconnections with Portugal in real life are numerous and of small size in some cases. Here they are represented with one interconnection in the north and one in the south of Portugal. The total interconnection capacity with each country is consistent with the aggregated values assumed in the Spanish NECP for 2030 ($5\,000$ MW with France and $4\,000$ with Portugal). Figure \ref{fig_interconnections} shows the interconnections included in the clustered network with 15 nodes.

\begin{table}[!ht]
\centering
\small
\begin{tabular}{cccc}
\hline
\textbf{Country} & \textbf{Name} & \textbf{Capacity [MW]} & \textbf{Length [km]} \\ 
\hline
France & ES FR0 & $2\,900$ & $34.25$ \\
France & ES FR1 & $2\,100$ & $200$ \\
Portugal & ES PT0 & $2\,000$ & $100$ \\
Portugal & ES PT1 & $2\,000$ & $100$ \\
\hline
\end{tabular}
\caption{Default parameters of the interconnections implemented in PyPSA-Spain. Length refers to the Spanish side, and does not necessarily correspond with the length between the border node and the closest one of the clustered network.}
\label{tab_interconnections}
\end{table}

\begin{figure}[!ht]  
\centering
\includegraphics[width=12cm]{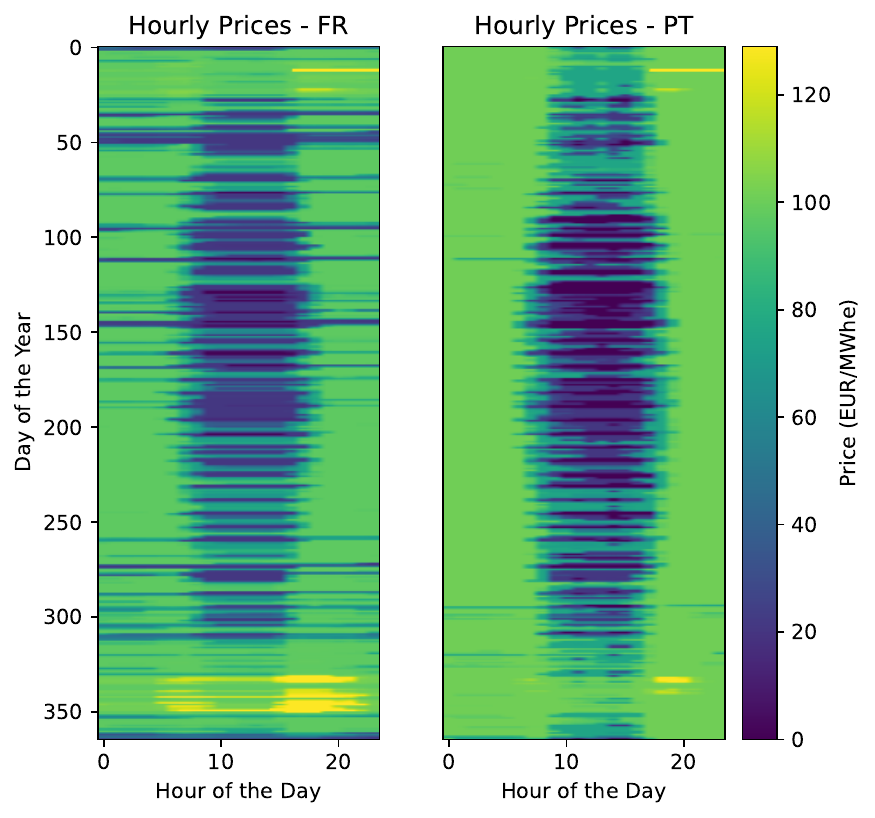}
\caption{Hourly electricity prices in France and Portugal obtained with a one-node per country optimisation with PyPSA-Eur.} \label{fig_prices}
\end{figure}

Finally, Figure \ref{fig_prices} shows the electricity prices for France and Portugal obtained with PyPSA-Eur, using one node per country and a global CO2 target of -70\% of CO2 emissions in the electricity sector, relative to 1990. This is a reasonable target for 2030, when the official global CO2 target in the UE is -55\% of emissions. The maximum value in the plot corresponds to the 99th percentile of prices in France, 128.53 EUR/MWh, but much higher prices are obtained for a few hours in winter. The figure shows a strong price slump during midday hours, due to the massive PV deployment in South Europe in an optimal decarbonised system. This is an important consideration to take into account when optimising the electricity mix in a particular country.

%% file: 03_s4_updated_esios.tex
\subsubsection{Improving the installed power capacity at regional level} \label{subsec_updated_esios}

PyPSA-Eur provides an initial distribution of power capacity at the high resolution scale according to several open databases. If these datasets do not cover solar and wind farms individually, a heuristic layout of the capacity proportional to the capacity factor of the Voronoi cells is performed for these technologies.

In Spain, the TSO provides data on the installed capacity of the different technologies at NUTS 2 level\footnote{\url{https://api.esios.ree.es/}}. The data for wind, solar PV, CCGT and nuclear in 2022 is included in PyPSA-Spain repository. These data are used to update the initial and exogenous distribution of power capacities established with PyPSA-Eur by increasing or decreasing the capacity of a given technology in the nodes within each NUTS 2 region to match the aggregated capacity in that region. Two options for increasing the capacity are implemented:
\begin{itemize}
    \item \textit{proportional}: the capacity in every node is increased proportionally to the initial capacity assigned in PyPSA-Eur, so that the aggregated capacity in the NUTS 2 region matches the historical data.
    \item \textit{additional}: the capacity is increased by the same amount. Only nodes with non-zero capacity are considered.
\end{itemize}

In both cases, it is ensured that the maximum capacity in a node due to land availability in the corresponding Voronoi cell (for solar and wind technologies) is not exceeded. If the capacity in the nodes within a NUTS 2 region needs to be reduced to match the historical data, the \textit{proportional} approach is used.

This functionality is particularly useful for comparing historical and modelled generation time series, as the same generation capacities of each technology are guaranteed at NUTS 2 level in both cases. It was employed to obtain the Q2Q transformations for wind power and solar PV generation in Section \ref{subsec_Q2Q}. This functionality is also useful to implement models with a certain power capacity deployment defined by exogenous scenarios (for example, those provided by policy-makers for 2030).

%% file: 03_s5_updated_gdp_pop.tex
\subsubsection{Updated \textit{gdp} and \textit{pop} weights} \label{subsec_updated_gdp_pop}

\begin{figure}[!ht]  \label{fig_gdp_pop_regressions}
\centering
\includegraphics[width=14cm]{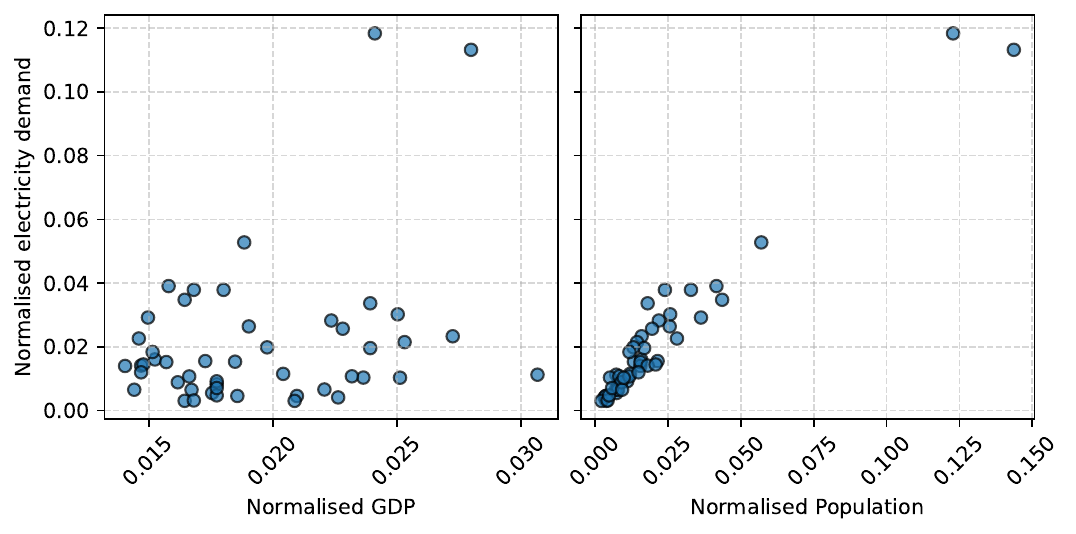}
\caption{Normalised electricity demand versus normalised GDP and population for Spanish NUTS 3 regions for year 2022.}
\end{figure}

As mentioned in Section \ref{sec_02_pypsa-eur}, in PyPSA-Eur the annual electricity demand of a country is distributed over the Voronoi cells in proportion to the GDP (as a proxy for industrial activity) and population of the NUTS 3 regions overlapped by the Voronoi cell. The weights used were calculated from a regression for European countries using country-level data, yielding \textit{gdp}$=0.60$ and \textit{pop}$=0.40$ (see \cite{Hoersch2018} for details). 
These weights may not fit well for single countries. Figure \ref{fig_gdp_pop_regressions} shows the electricity demand in NUTS 3 regions in Spain in 2022 versus GDP (left) and population (right). It can be seen that, in this case, population has a higher correlation with electricity demand than GDP. 

A linear regression in the form: 

$$ \frac{demand_{i}}{demand_T} = gdp \frac{GDP_{i}}{GDP_T} + pop \frac{population_{i}}{population_T}, $$

\noindent where the sub-index $i$ refers to the NUTS 3 region and $T$ is the total for Spain, yielded $$gdp=0.16 ~~~,~~~ pop=0.82.$$ 

These weights are employed in PyPSA-Spain, replacing those used in PyPSA-Eur.

%% file: 03_s6_costs.tex
\subsubsection{Including technology cost assumptions}  \label{subsec_costs}

Economic data on the technologies included in PyPSA-Eur are retrieved from various sources, mostly from the Danish Energy Agency (DEA)\footnote{See \url{https://github.com/pypsa/technology-data}}. These data are particularised for the target year \texttt{yyyy} and collected in file \texttt{resources/costs\_yyyy.csv}. The same data are used for all the countries considered in the model, disregarding differences between countries due to, for example, labour costs, taxes or capital costs. 

In PyPSA-Spain, modifications in technology costs can be defined in the file 

\noindent \texttt{data\_ES/costs/costs\_updates.csv}. Each row in this file is in the form of the triple \textit{(technology, parameter, value)}. This functionality allows providing the model with a refined estimation of the costs for the particular case of Spain.

%% file: 04_impacts.tex
\section{Impact of the new functionalities on the optimal mix} \label{sec_04_impacts}

This section analyses the impact of the main functionalities described in Section \ref{sec_03_pypsa-spain}. The analysis is carried out by comparing the optimal configurations of the Spanish power system  obtained for the following cases:

\begin{itemize}
    \item \textbf{Reference case}: obtained with PyPSA-Eur considering only Spain (no interconnections included).
    \item \textbf{Case 1}: including only the improved estimation of renewable profiles, as described in Section \ref{subsec_Q2Q}.
    \item \textbf{Case 2}: including only the improved characterisation of the electricity demand, as described in Section \ref{subsec_load_dist}.    
    \item \textbf{Case 3}: including only the interconnections, as described in Section \ref{subsec_interconnections}.
\end{itemize}

The following characteristics are common to all the cases:

\begin{itemize}

    \item The annual electricity demand is 344 TWh. This corresponds to the electricity demand for peninsular Spain in 2030 assumed in the Spanish National Energy and Climate Plan (NCEP, or PNIEC by its Spanish acronym) \cite{MITERD2024}.

    \item The CO2 limit for the electricity sector is set at 11.986 Mt, in line with the emissions associated with the peninsular power system in the Spanish NECP for 2030. To provide context, the CO2 emissions from the electricity sector in peninsular Spain were of around 25 Mt in 2023, according to REE, the Spanish TSO, and of around 65.575 Mt in 1990.\footnote{\url{https://www.ree.es/es/datos/generacion/no-renovables-detalle-emisiones-CO2}} Hence, the CO2 target corresponds to 18.3\% of 1990 emissions.

    \item The generation capacity to be optimised in this study includes onshore wind, floating offshore wind, solar PV and CCGT. For storage capacity, batteries and H2 storage (steel tank and underground) were included.
    
    \item Exogenous fixed capacities have been assumed for certain technologies, see Table \ref{tab_non_ext_capacities}. For nuclear power, the capacity corresponds to the remaining capacity in 2030 according to the Spanish nuclear decommissioning plan.
        
    \begin{table}[!ht]
    \centering
    \small
    \begin{tabular}{lr}
    \hline
    \textbf{Technology} & \textbf{Capacity [MW]}\\ 
    \hline
    Hydro (reservoir) & $14\,965$  \\
    Hydro (pumped storage) & $8\,870$  \\
    Hydro (run-of-river) & $277$  \\    
    Nuclear & $3\,112$  \\
    Biomass &  $533$ \\
    \hline
    \end{tabular}
    \caption{Fixed exogenous capacities assumed for 2030, see text for details.}
    \label{tab_non_ext_capacities}
    \end{table}

    \item Several numbers of clusters were considered to assess the effect of the spatial resolution on the results, from 2 (peninsular Spain and Balearic islands) to 100 clusters.

    \item The assumed investment costs are summarised in Table \ref{tab_costs}. They correspond to the values provided for 2030 in the repository technology-data\footnote{\url{https://github.com/PyPSA/technology-data}} (v0.9.1), except for solar PV where 440 EUR/kWe is assumed here. This value is a weighted average resulting from taking into account 75\% of solar utility at 380 EUR/kWe and 25\% of solar rooftop at 620 EUR/kWe. This share is in line with the Spanish NECP, which assumes 19 GW of distributed generation out of 76 GW of solar PV.

    \begin{table}[!ht]
    \centering
    \small
    \begin{tabular}{lrl}
    \hline
    \textbf{Technology} & \textbf{Investment cost} & \textbf{Units} \\ 
    \hline
    Solar PV & $440.0$ & EUR/kW$_\text{e}$ \\ 
    Wind power onshore & $1\,095.8$ & EUR/kW$_\text{e}$ \\
    Wind power floating offshore & $2\,350.0$  & EUR/kW$_\text{e}$ \\ 
    Battery (inverter) & $169.3$ & EUR/kW$_\text{e}$ \\
    Battery (storage) & $150.3$ & EUR/kWh \\    
    \hline
    \end{tabular}
    \caption{Investment cost assumptions for 2030. See text for details.}
    \label{tab_costs}
    \end{table}
    
    \item As a conservative approach, the lifetime of all renewable technologies was assumed to be 20 years.

    \item The renewable capacity per squared kilometre considered for all the renewable technologies (solar PV, onshore wind and floating offshore wind) was set at 1 MW/km$^2$. Actual power densities achieved in wind farms and solar PV installations may be 5 to 100 times higher. The reason for this conservative choice is to limit the expansion of renewables to a percentage of available land, which is a more sensitive approach to public acceptance issues.

\end{itemize}

\

\textbf{Results for the Reference case}

\begin{figure}[!ht]  
\centering
\includegraphics[width=14cm]{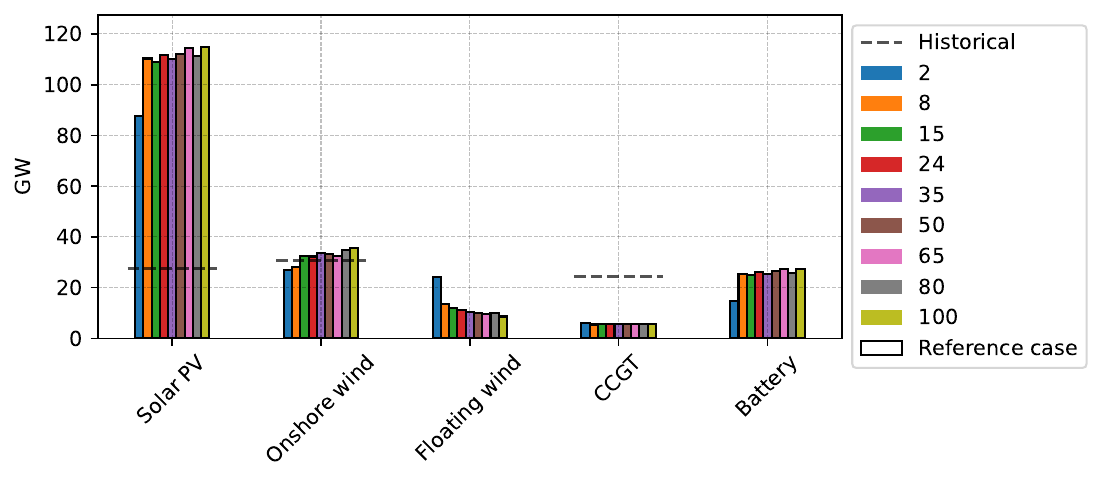}
\caption{Optimal mix for the Reference case.} \label{capacities_00_updated_reference}
\end{figure}

Figure \ref{capacities_00_updated_reference} shows the optimised capacities for the Reference case. For each technology, the set of bars corresponds to a different number of clusters of the modelled network. To provide context, the horizontal dashed lines represent the actual capacities in 2024 for peninsular Spain for solar PV, onshore wind power and CCGT.

The results show a power system heavily based on solar PV, about four times the actual capacity, and about 25 GW of newly installed batteries (with around 150 GWh of energy capacity, not shown in the figure). Battery requirements could be lower in a scenario that considers sector coupling with electric vehicles, where vehicle batteries could be used in part to balance the system. Also noteworthy is the installation of around 10 GW of floating offshore wind turbines, while no new onshore wind capacity is included. The optimal capacity for CCGTs is significantly lower than the current capacity, which is consistent with the current situation of overcapacity (equivalent hours for CCGTs in the last decade in Spain are in the range of 500-2000). The impact of the selected number of nodes on the optimal capacities is low, but not negligeable. In particular, higher number of nodes leads to an increase in onshore wind capacity at the expense of floating offshore. This is because higher spatial resolutions allow the identification of high potential onshore wind sites. Finally, it is noted that no hydrogen storage appears in the optimal configuration. This can be explained by two facts: first, the CO2 limit is not very strit, allowing CCGTs to generate in situations with low wind or solar resource. Second, only the electricity demand is considered in the model, and hydrogen storage is characterised by a low round-trip efficiency. Including hydrogen demand through sector coupling would probably lead to relevant amounts of electrolysers and fuel cells in the optimal configuration.

\

\textbf{Results for Case 1}

\begin{figure}[!ht]  
\centering
\includegraphics[width=14cm]{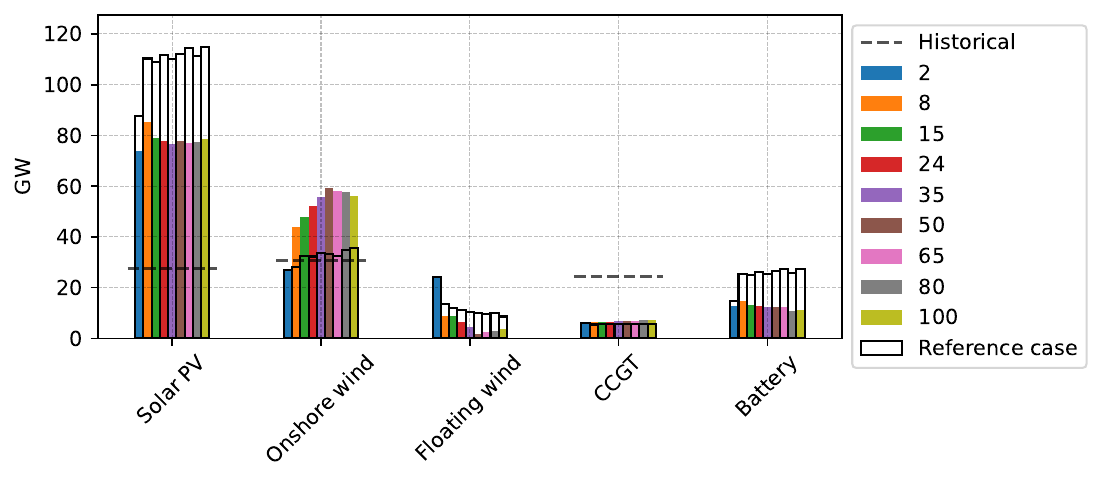}
\caption{Optimal configuration for Case 1.} \label{capacities_01_updated_q2q}
\end{figure}

Figure \ref{capacities_01_updated_q2q} shows the optimal capacities when the Q2Q transformations obtained in Section \ref{subsec_Q2Q} are considered. 

The improved estimation of the hourly capacity factors leads to a clearly different optimal mix as compared with the Reference case. It is characterised by a more balanced share of solar PV and wind, and less needs for batteries ($\sim 10$ GW) and floating offshore ($\sim 5$ GW). This result was to be expected, as the Q2Q transformation allowed the underestimation of the capacity factor of onshore wind to be corrected, see discussion in Section \ref{subsec_Q2Q}.

\

\textbf{Results for Case 2}

\begin{figure}[!ht]  
\centering
\includegraphics[width=14cm]{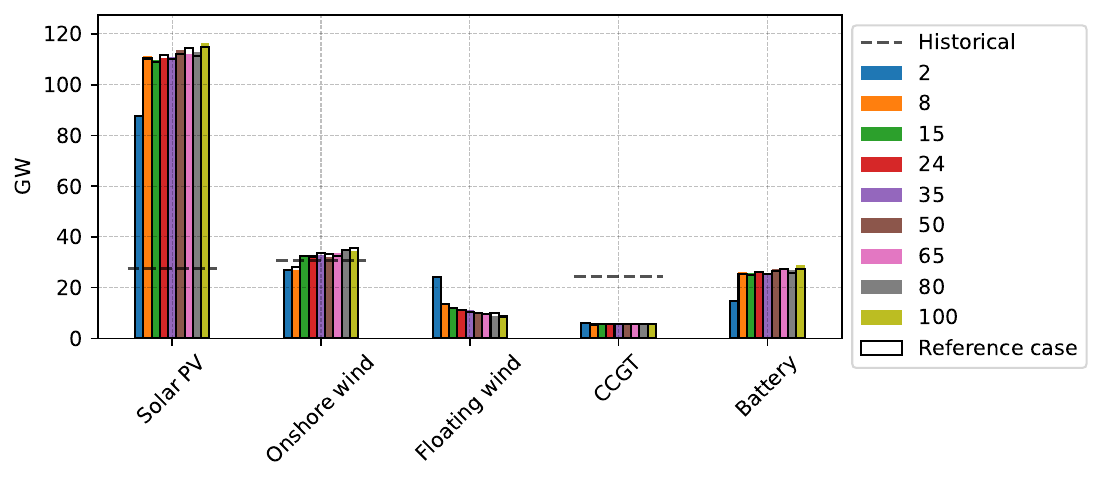}
\caption{Optimal configuration for Case 2.} \label{capacities_02_updated_load}
\end{figure}

Figure \ref{capacities_02_updated_load} shows the results when the spatio-temporal distribution of electricity demand was considered (see Section \ref{subsec_load_dist}.

In this case, the variation of optimal capacities with respect to the Reference case is almost negligible. This suggests that the spatio-temporal features of the electricity demand captured by the implemented methodology could be handled through the transmission lines, back-up generation (CCGT and hydro) and battery storage, without relevant changes in optimal aggregated capacities, at least for the 2030 decarbonistaion target. Further analyses considering more ambitious CO2 targets are required to assess the relevance of the improved spatio-temporal electricity demand distribution on the optimal mix.

\

\textbf{Results for Case 3}

\begin{figure}[!ht]  
\centering
\includegraphics[width=14cm]{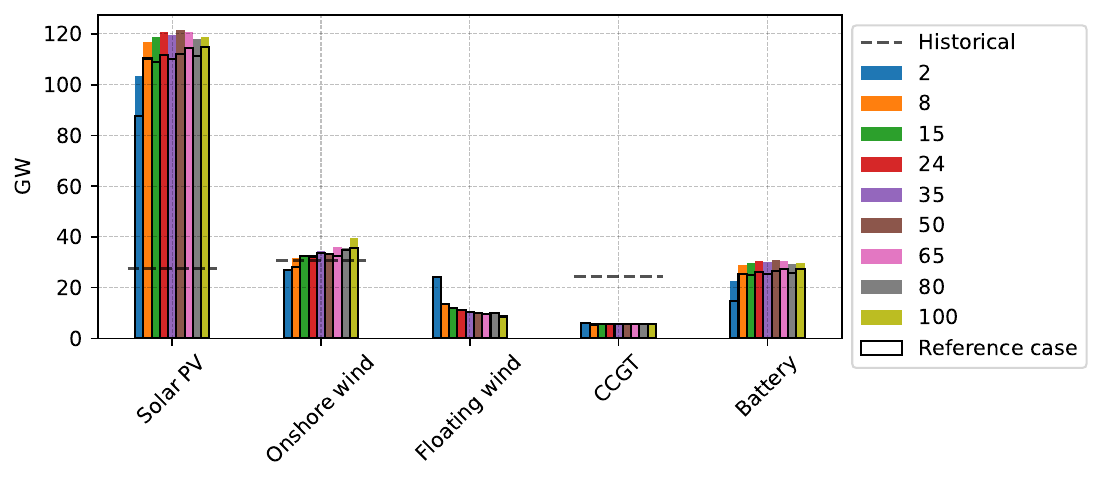}
\caption{Optimal configuration for Case 3.}
\label{capacities_03_updated_ic}
\end{figure}

Figure \ref{capacities_03_updated_ic} shows the optimal mix when the interconnections with France and Portugal were considered, as described in Section \ref{subsec_interconnections}. 

\begin{figure}[!ht]  
\centering
\includegraphics[width=14cm]{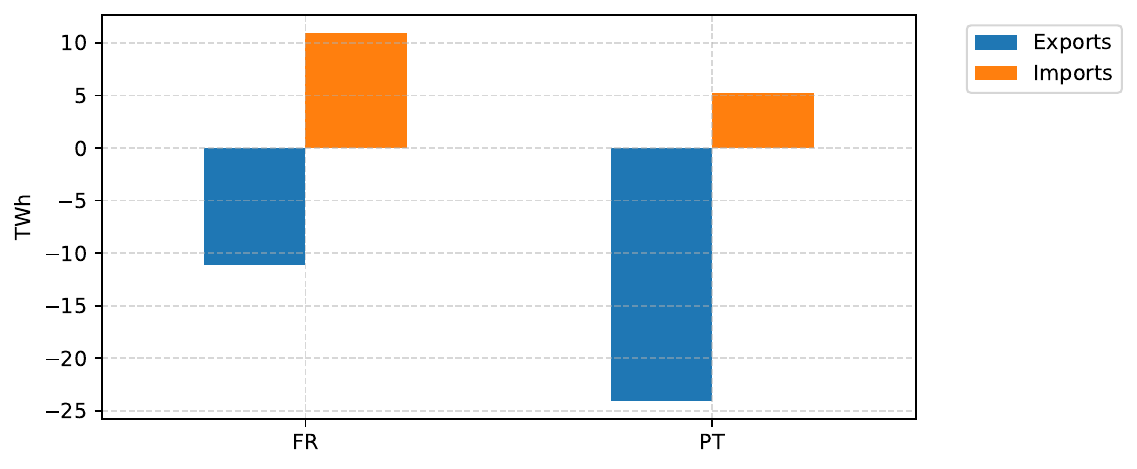}
\caption{Annual electricity exchanges through interconnections.}  \label{imports_exports_50nodes}
\end{figure}

It can be observed that the installed capacities in this case do not change much as compared to the Reference case (without interconnections), only an increase of $\sim 8\%$ for solar PV and $\sim 15\%$ for batteries. However, it is interesting to analyse the annual electricity exchanges for each interconnection, as it is shown in Figure \ref{imports_exports_50nodes} for the case with 50 clusters. According to this, Spain would be a net exporter to France and Portugal with 0.7 TWh/a and 18.8 TWh/a of net exports respectively. To provide context, according to the Spanish TSO,\footnote{\url{https://www.sistemaelectrico-ree.es/informe-del-sistema-electrico/intercambios/resumen-intercambios-internacionales}} in 2023 the balance with France was a net export of 1.8 TWh, and 9.9 TWh with Portugal, which shows that the obtained results are of the same order of magnitude. However, it is important to note that the net import/export character with each country may vary according to the interplay between the different hypotheses assumed in the considered scenario. 

%% file: 05_PNIEC.tex
\section{Comparison with the Spanish NECP} \label{sec_05_PNIEC}

In this section, the optimal mix for 2030 is obtained with PyPSA-Spain, considering all the new functionalities implemented. The results are compared with the Spanish National Energy and Climate Plan (NECP, or PNIEC by its Spanish acronym) \cite{MITERD2024}.

The Spanish NCEP sets out the Spanish climate and energy targets required of all member states by the European Commission to support the European Union's commitments under the Paris Agreement. In particular, the document sets out a roadmap for Spain to achieve its 2030 targets and lay the foundations for long-term decarbonisation by 2050. A first version of the Spanish NECP was delivered in January 2020. A revised draft was published in June 2023, reflecting an increase in ambition, mainly due to the energy crisis in Europe followed by the Russia's invasion of Ukraine. The final update of the Spanish NECP was published in September 2024, and is the document employed for discussion here. 

The power system in the Spanish NECP was modelled with PLEXOS, a commercial software for power system optimisation. In the appendix D of the Spanish NCEP, a detailed description of the hypotheses assumed is included. Two of them are here remarked, as they represent two key aspects that can be overcome with PyPSA-Eur:

\begin{itemize}
    \item The employed model assumes the copper-plate hypothesis (\textit{i.e.} infinite capacity of the transmission grid).
    \item The installed capacity mixes for the different European countries are the current ones.
\end{itemize}

The first hypothesis makes it impossible to identify grid bottlenecks associated with new capacity deployment, or to calibrate the grid expansion requirements associated with the proposed optimal mix. This is overcome in PyPSA-Eur by considering the spatial resolution, which allows to capture the trade-off between exploiting sites with high renewable resources and the associated grid expansion costs.

The second hypothesis means that no energy transition is assumed in the neighbouring countries. This may lead to an inaccurate characterisation of electricity exchanges between countries and, consequently, to a poorly calibrated electricity mix.

\begin{figure}[!ht]  
\centering
\includegraphics[width=14cm]{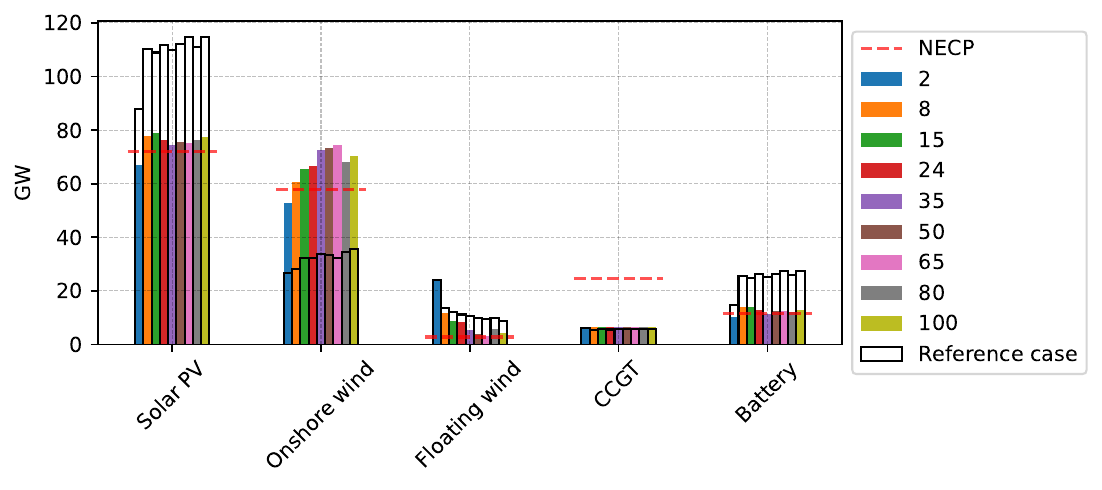}
\caption{Optimal configuration for 2030 estimated with PyPSA-Spain. Horizontal lines correspond to the expected capacities according to the Spanish NECP.} \label{capacities_updated_04_ALL}
\end{figure}

Figure \ref{capacities_updated_04_ALL} shows the optimal capacities according to technology and number of clusters obtained with PyPSA-Spain. The Reference case, as in the previous section, corresponds to the capacities obtained with PyPSA-Eur for Spain only (isolated). In the figure, the horizontal lines correspond to the expected capacities in 2030 according to the Spanish NECP. For batteries, the line corresponds to the difference between the 17.6 GW of stores expected in 2030 minus the 6 GW of pumped hydro storage existing in 2020.

In general, the results obtained with PyPSA-Eur are in good agreement with the Spanish NECP. A more balanced generation mix between onshore wind and solar PV is here obtained (around 70 GW each). The impact of the spatial resolution on the optimal capacities is noticeable for wind (onshore and offshore), where the optimal capacities stabilise for 35 nodes and above. This suggests a minimum spatial resolution to be considered in future studies using PyPSA-Spain.

\begin{figure}[!ht]  
\centering
\includegraphics[width=7.8cm]{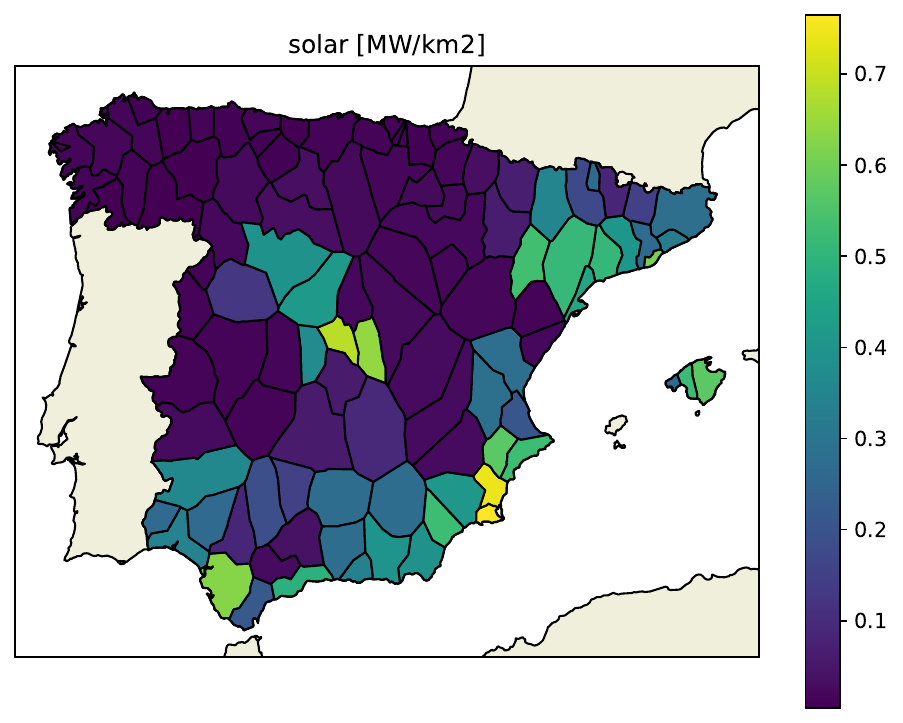}
\includegraphics[width=7.8cm]{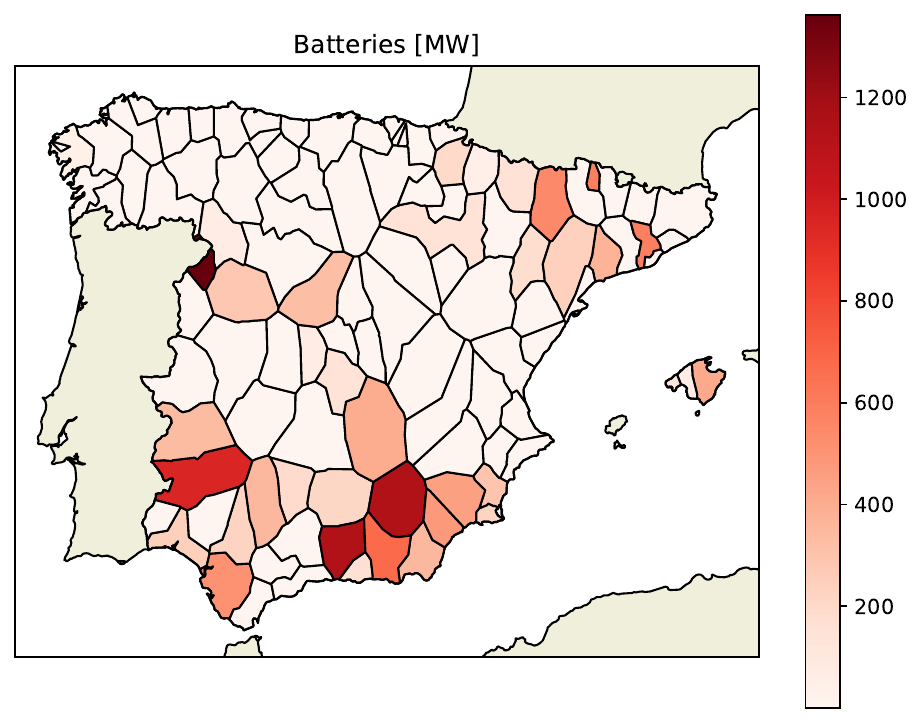}
\caption{Optimal geographical capacity distribution for solar PV (left) and batteries (right).} \label{PNIEC_solar_batt}
\end{figure}

Figure \ref{PNIEC_solar_batt} (left) shows the geographical distribution of optimal solar PV in terms of capacity density (MW/km$^2$) for the highest spatial resolution considered (100 nodes). High solar PV capacities can be observed in the south of Spain, where the resource is high, and also in central and northeastern Spain, close to large metropolitan areas (Madrid and Barcelona), where electricity demand is high. This result underlines the importance of installing solar PV close to areas of high electricity demand. Note that the high demand densities identified in Figure \ref{fig_annual_demand} (left) in north-central Spain do not imply large PV capacity, probably due to the low solar resource as compared with the wind resource, and to the fact that the demand in this region is mainly industrial, which correlates worse with the PV generation profile (see Appendix \ref{app_load}). 

Figure \ref{PNIEC_solar_batt} (right) shows the optimal distribution of batteries, which correlates relatively well with PV capacity in high resource areas (South Spain). Additionally, about 1 GW of batteries are also installed near the two interconnections with Portugal. An analysis of the electricity flows at these nodes showed that the main purpose of these batteries is to store imports from Portugal, which occur mainly during the summer with a daily pattern, when there is likely to be an excess of solar PV generation also in Portugal.

\begin{figure}[!ht]  
\centering
\includegraphics[width=7.8cm]{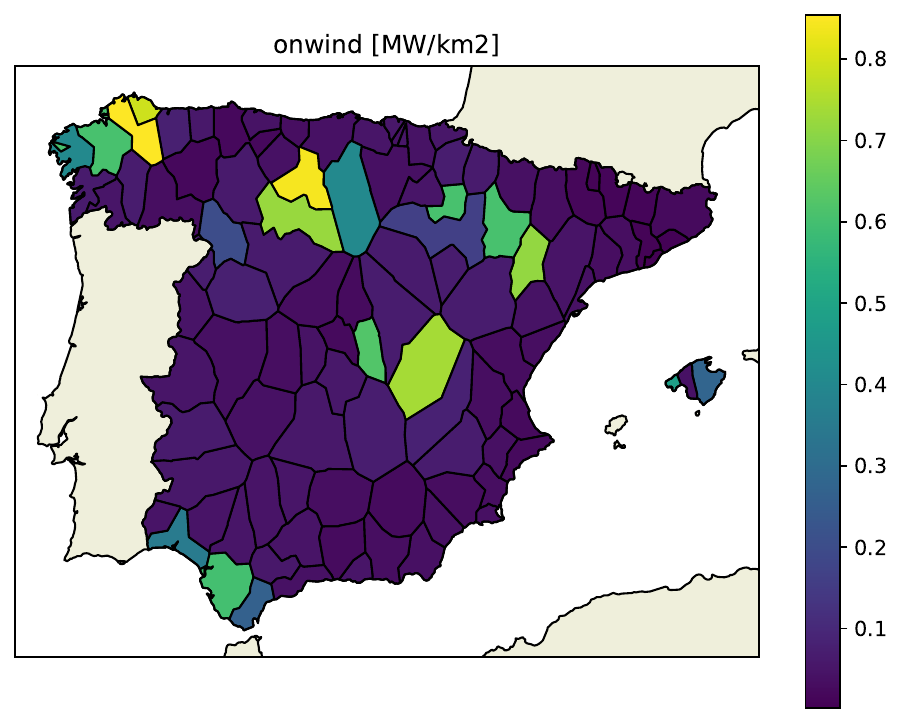}
\includegraphics[width=7.8cm]{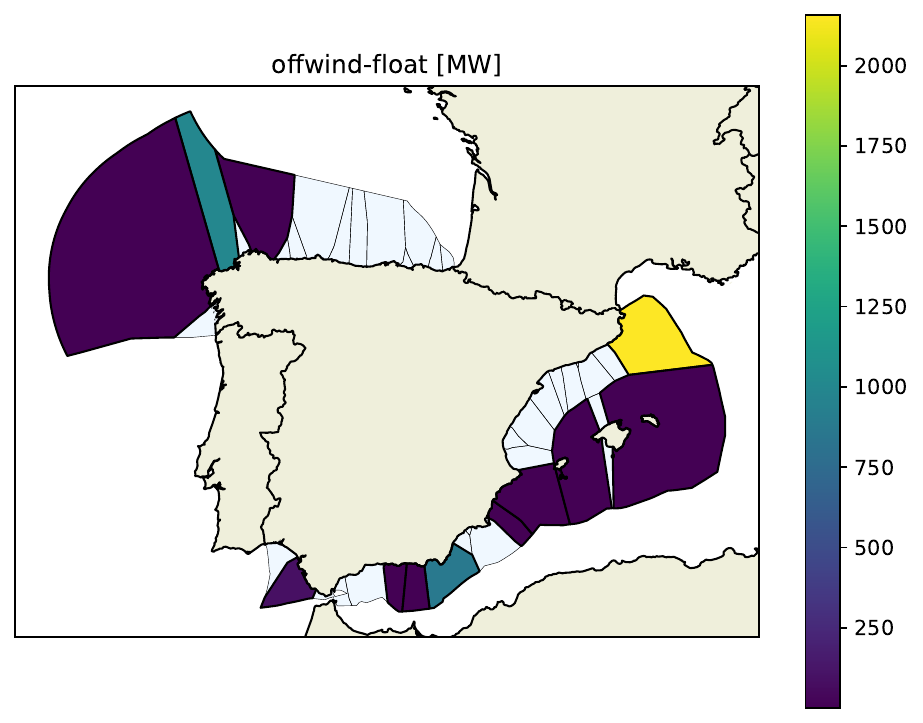}
\caption{Optimal geographical capacity distribution for onshore wind (left) and floating offshore (right).} \label{PNIEC_onshore_offshore}
\end{figure}

Figure \ref{PNIEC_onshore_offshore} shows the optimal distribution of onshore (left) and offshore (right) wind power capacity (onshore capacity in terms of capacity density, MW/km$^2$). For the onshore case, the local annual capacity factor seems to be the main driver of the resulting capacity layout, see Figure \ref{fig_scheme}-c).

Appendix \ref{app_PNIEC_NUTS3} includes figures with the share of the optimal mix for different technologies by Spanish provinces (NUTS 3 level). They were estimated from the highest resolution results (100 clusters), and represent an interesting outcome for policymakers, made possible by the spatial resolution of the model.

The annual capacity factor of each generation technology is compared to those included in the Spanish NECP in Table \ref{tab_CF}. A relatively good agreement is observed, except for CCGT. The discrepancy is due to the fact that the Spanish NECP considers the actual CCGT capacity (24.5 GW), while the results from PyPSA-Eur show a much lower optimal capacity required (around 6 GW).

\begin{table}[!ht]
\centering
\begin{tabular}{lcc}
\hline
\textbf{Technology} & \textbf{PyPSA-Spain}  & \textbf{NECP}\\ 
\hline
Solar PV            & 0.177  & 0.198 \\
Onshore wind power  & 0.264  & 0.232 \\
Floating offshore wind power   & 0.476 & 0.395  \\
Nuclear          & 0.810  & 0.825\\
CCGT             & 0.619  & 0.092 \\
\hline
\end{tabular}
\caption{Annual CF for generation technologies} \label{tab_CF}
\end{table}

\begin{figure}[!ht]  
\centering
\includegraphics[width=14cm]{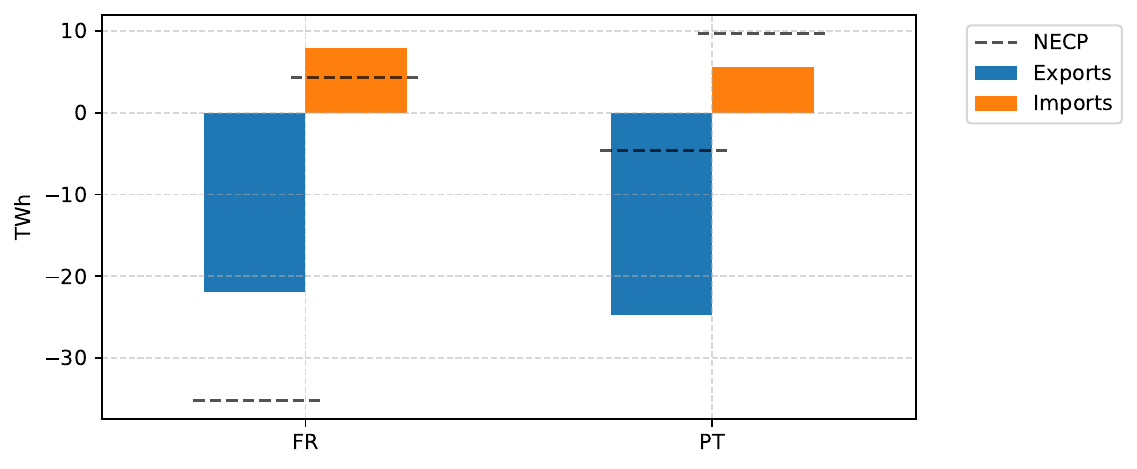}
\caption{Annual electricity exchanges through the interconnections.} \label{imports_exports_100nodes}
\end{figure}

In terms of annual electricity exchanges through interconnections, Figure \ref{imports_exports_100nodes} shows a situation where Spain is a net exporter to both France (with a net export of 14.0 TWh/a) and Portugal (19.1 TWh/a).
According to the Spanish NECP scenario, Spain would export 31.0 TWh/a to France and import 5.0 TWh/a from Portugal. The notable discrepancies could be explained by the fact that, as mentioned above, the Spanish NECP assumes actual capacity mixes for neighbouring countries. This is likely to result for example in systematic exports from Spain to France during the midday hours due to the high expected installed PV capacity in Spain. However, PyPSA-Spain considers a scenario for 2030 that includes decarbonisation across Europe, in which solar PV capacity would also increase significantly in France. This suggests that the high exports from Spain to France reflected in the Spanish NECP could be overestimated.

\begin{figure}[!ht]  
\centering
\includegraphics[width=14cm]{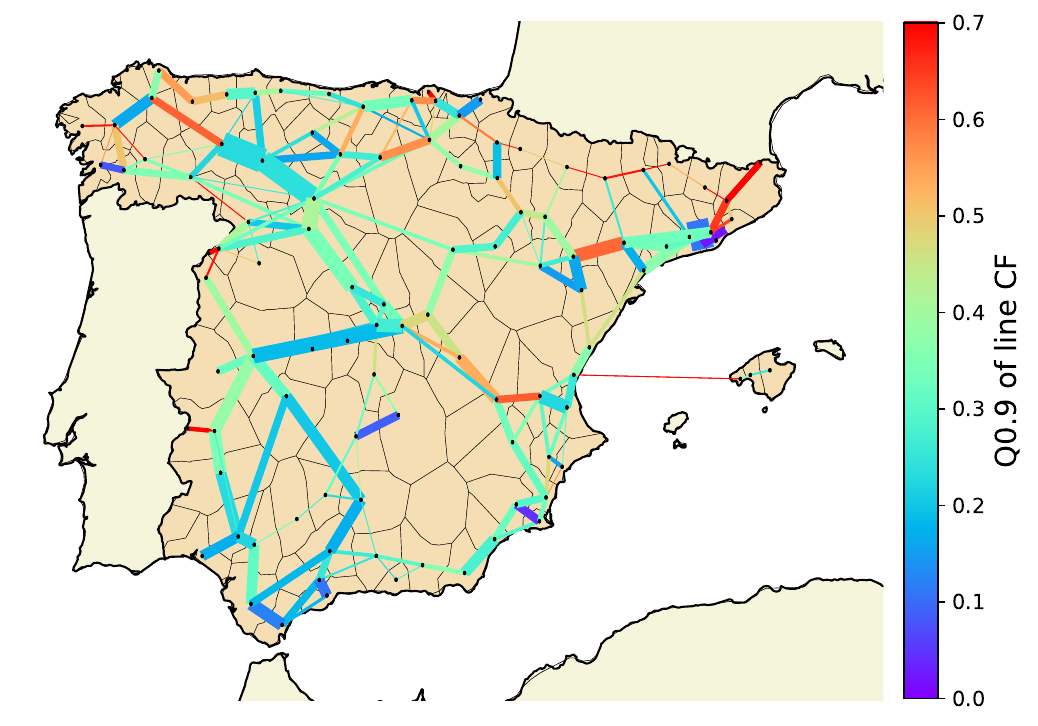}
\caption{Percentile 90 of the hourly line loading for the network model with 100 nodes.} \label{fig_bottlenecks}
\end{figure}

Concerning the grid requirements, this analysis did not consider the possibility of grid expansion. However, potential bottlenecks can be identified from the optimal dispatch. Figure \ref{fig_bottlenecks} shows the 90 percentile of the hourly line loading for the network with 100 nodes. Note that the maximum flow on each line has been exogenously set to 0.7 of its capacity to consider the N-1 security criterion (except for interconnections). Thus, the lines in red are saturated for at least 10\% of the year. The figure shows regions that are poorly interconnected, such as the north-east (Catalonia), where demand is high, and the north-west (Galicia), where wind capacity is high.

Finally, it is worth mentioning that the CO2 price obtained with PyPSA-Spain is in the range of $71.2-77.1$ EUR/tCO2 for the different spatial resolutions. This price results from the CO2 constraint defined in the optimisation problem, and reflects the required carbon price that makes the obtained generation mix the most cost-effective solution of those compatible with the CO2 limit. In the Spanish NECP, the carbon price was set exogenously at 79 EUR/tCO2.

%% file: 06_conclusions.tex
\section{Conclusions} \label{06_conclusions}

This paper presents PyPSA-Spain, an open-source high-spatial resolution model for the Spanish energy system based on PyPSA-Eur. PyPSA-Spain is an up-to-date fork of PyPSA-Eur, ensuring that new functionalities and bug fixes made to PyPSA-Eur are integrated.

The primary motivation behind the development of PyPSA-Spain was to leverage the benefits of a national energy model over a European one, like the availability of specific datasets from national organisations. Additionally, a single-country model enables higher spatial and temporal resolution with the same computational resources due to the smaller geographical domain. Finally, it does not require assumptions about coordinated action between countries, making it a more suitable tool for analysing national energy policies. To accommodate cross-border interactions, a nested model approach with PyPSA-Eur was used, wherein electricity prices from neighbouring countries are precomputed through the optimisation of the European energy system for the same planning horizon and weather year.

PyPSA-Spain includes a number of novel functionalities that enhance the representation of the Spanish energy system, as compared with PyPSA-Eur. The main ones are: (i) improved estimation of hourly capacity factors for solar PV and onshore wind, calibrated and validated with historical data; (ii) improved spatio-temporal representation of the electricity demand, based on historical electricity demand profiles at both NUTS 2 and NUTS 3 level; (iii) a flexible model for cross-border interconnections, allowing the optimisation of electricity exchanges based on price differences between Spain and neighbouring countries.

These new functionalities were evaluated for their impact on the optimal energy mix. The greatest influence was observed from the enhanced renewable profiles, leading to a recalibrated balance between solar PV and wind power capacities. In contrast, the refined electricity demand representation had minimal effect on the optimal energy mix, though the methodology remains valuable for modelling future demand scenarios with varying residential, services and industry shares. A sensitivity analysis also showed the need to consider a high spatial resolution of at least 35-50 nodes, especially to fully capture the onshore wind potential.

Finally, PyPSA-Spain has been applied to determine optimal configurations for achieving the decarbonisation goals for 2030 outlined in the Spanish National Energy and Climate Plan (NECP, or PNIEC by its Spanish acronym), revealing qualitatively similar renewable capacity requirements and CO2 price. The spatial resolution of the model allowed for a geographical description of the optimal renewable and storage capacity, revealing the importance of installing solar PV close to high demand areas when the demand has a marked daily profile (residential and services, not industrial). In addition, it was found that assumptions on the decarbonisation pathways followed by neighbouring countries may have an impact on the estimated electricity imports/exports. On this basis, a potential overestimation of electricity exports from Spain to France was identified in the Spanish NECP.

This first version of PyPSA-Spain focuses exclusively on the power system. Future research will cover other features included in PyPSA-Eur, such as sector coupling and multi-horizon modelling.

%% file: 07_others.tex
\section*{Acknowledgements}

This research was funded by the Spanish Ministerio de Universidades, through the national mobility program of the \textit{Plan Estatal de Investigación Científica, Técnica y de Innovación 2021-2023}, grant number CAS22/00069.

\section*{Code availability}

The code of PyPSA-Spain is available at \url{https://github.com/cristobal-GC/pypsa-spain}, and related documentation can be found at \url{https://pypsa-spain.readthedocs.io/en/latest/}. Most of the figures in this paper were generated with pypsa-Xplore, an open set of Jupyter notebooks for exploring the objects generated with PyPSA-Spain and PyPSA-Eur (\url{https://github.com/cristobal-GC/pypsa-Xplore}).

%% file: 99_app_Spain_Geospatial.tex
\section{Geospatial information of Spain} \label{app_Spain_Geospatial}

\begin{figure}[h]  
\centering
\includegraphics[width=7.5cm]{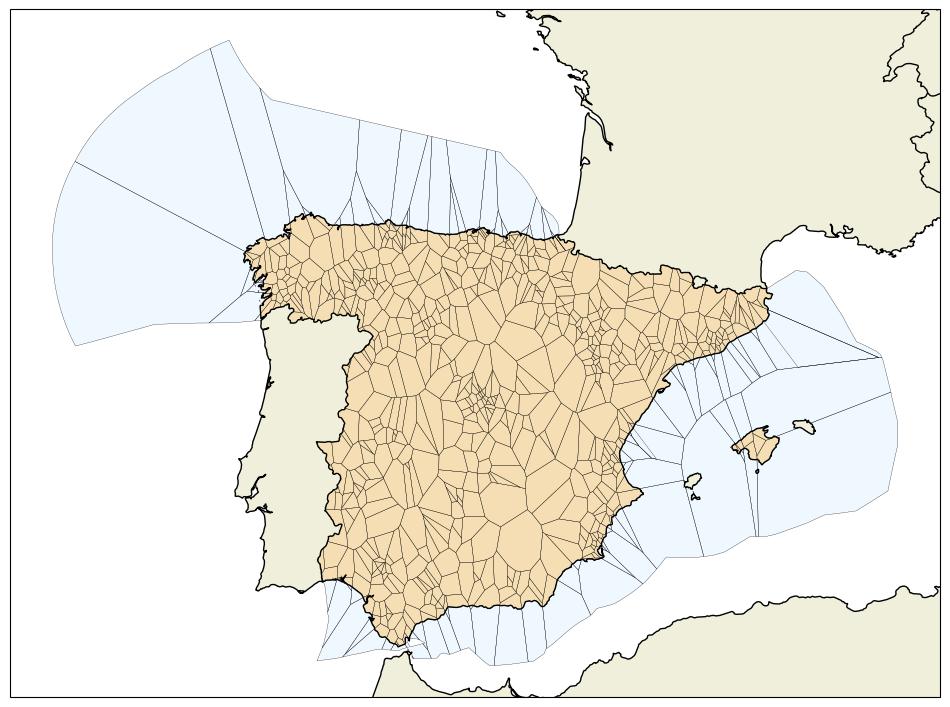}
\includegraphics[width=7.5cm]{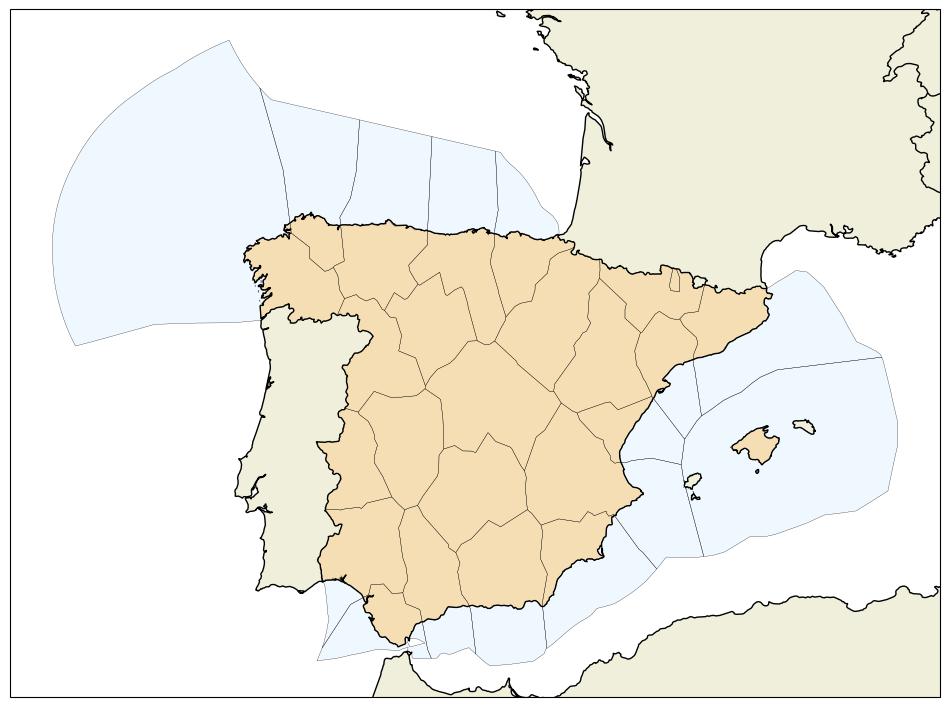}
\includegraphics[width=7.5cm]{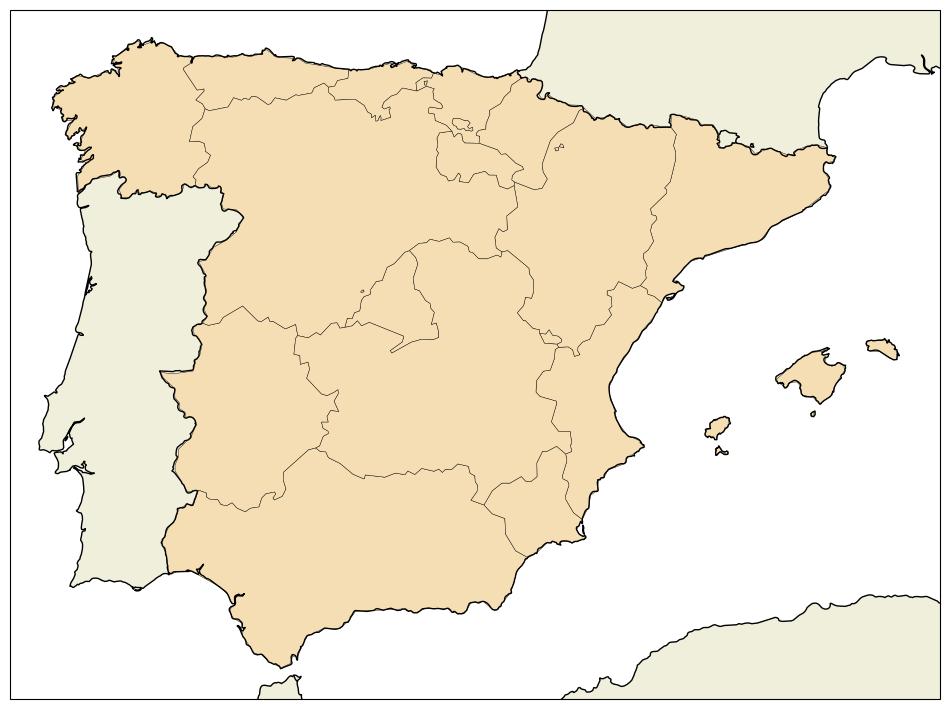}
\includegraphics[width=7.5cm]{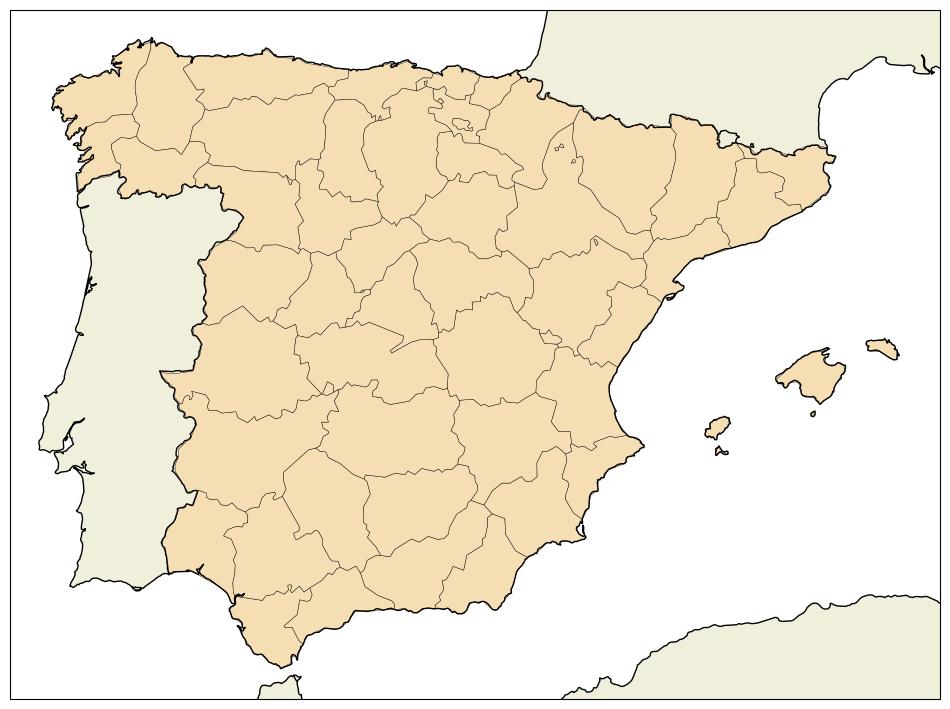}
\caption{Spatial subdivisions of Spain referred in this paper. Top left: Voronoi cells in the high-resolution network model; Top right: Voronoi cells for a network model with 24 nodes; Bottom left: NUTS 2 (autonomous communities); Bottom right: NUTS 3 (provinces).} \label{fig_Spain_Geospatial}
\end{figure}

Figure \ref{fig_Spain_Geospatial} shows the different spatial subdivisions of Spain referred in this paper. On top left, the Voronoi cells resulting from the high-resolution network ($\sim 700$). On top right, the resulting regions after clustering the network with 24 nodes. At the bottom left, 16 out of the 19 NUTS 2 regions (autonomous communities) in Spain.\footnote{Ceuta, Melilla and Canary Islands are excluded since they are not connected to the network modelled in PyPSA-Spain.} At the bottom right, 48 out of the 52 NUTS 3 regions (provinces) in Spain.\footnote{Ceuta, Melilla, Las Palmas and Santa Cruz de Tenerife excluded.}

\begin{figure}[!ht]  
\centering
\includegraphics[width=13cm]{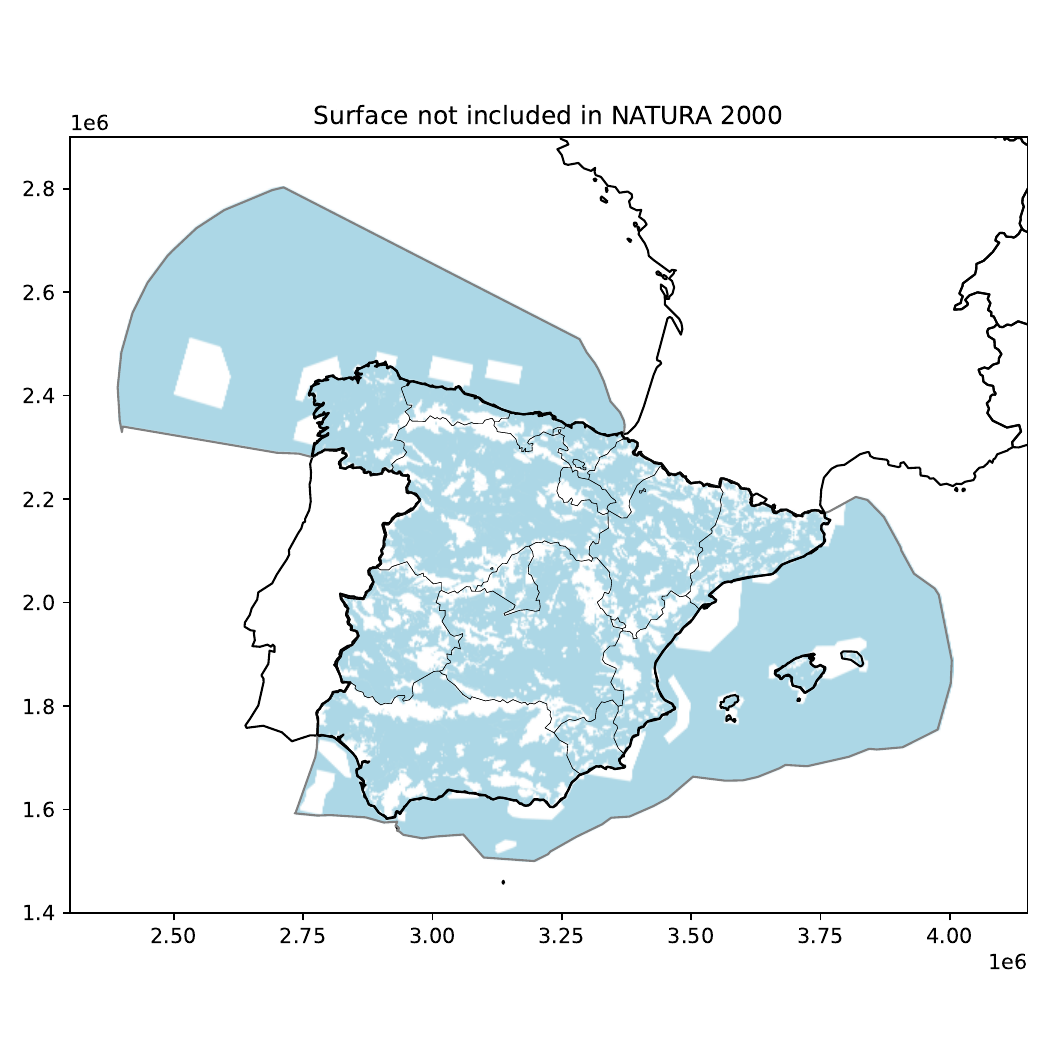}
\caption{Surface not protected by Natura 2000 network.} \label{Natura2000}
\end{figure}

Figure \ref{Natura2000} shows the surface in Spain (onshore and offshore) not protected by the Natura 2000 network. 

\begin{figure}[!ht]  
\centering
\includegraphics[width=10.05cm]{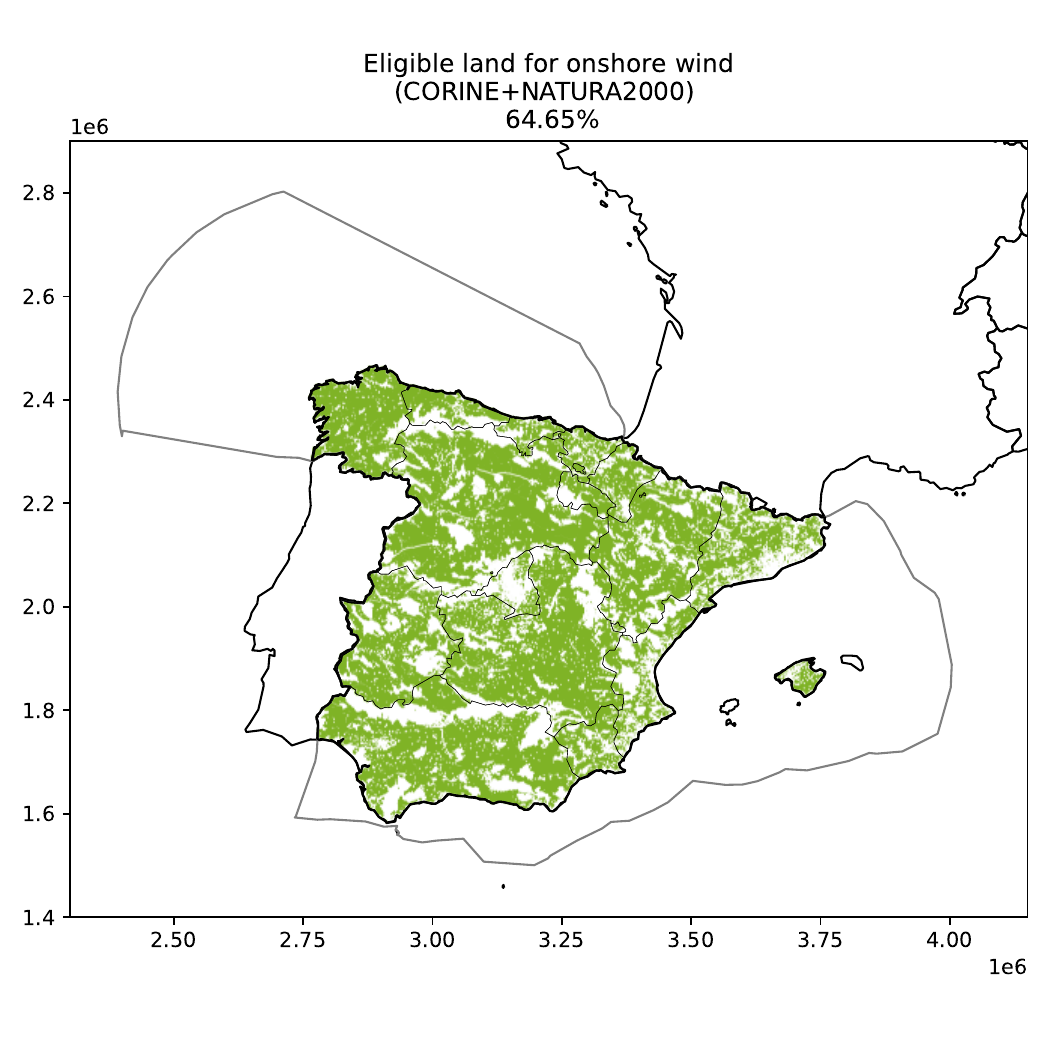}
\includegraphics[width=10.05cm]{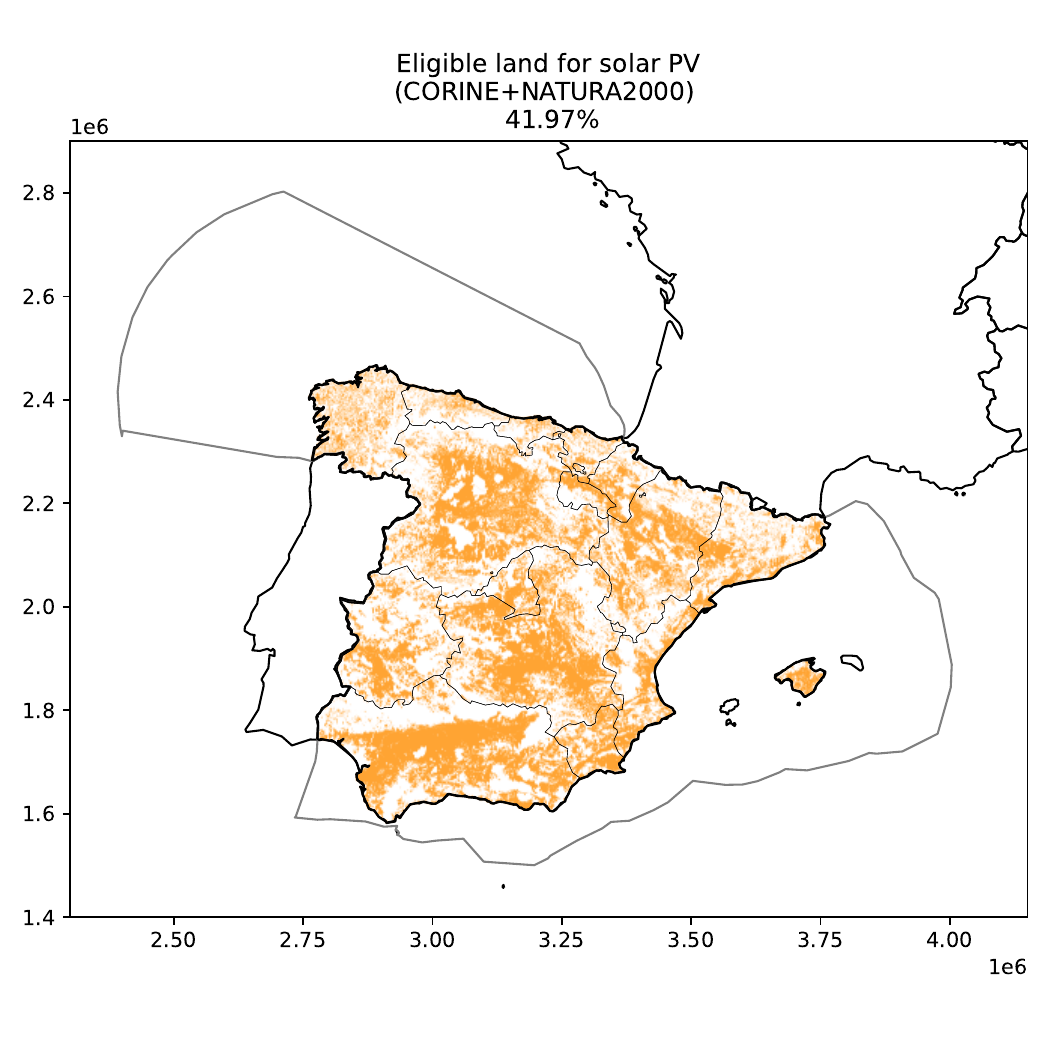}
\caption{Eligible terrain for onshore wind (top) and solar PV (bottom).} \label{CORINE_NATURA_2000}
\end{figure}

Concerning the CORINE land cover classification, the following codes were considered to define eligible terrain for onshore wind power installation: 
12, 13, 14, 15, 16, 17, 18, 19, 20, 21, 22, 23, 24, 25, 26, 27, 28, 29, 31, 32.
Additionally, the following codes were excluded with a buffer of 1 km: 1, 2, 3, 4, 5, 6. The resulting eligible terrain for onshore wind after combining the Natura 2000 layer and the CORINE layer is shown in Figure \ref{CORINE_NATURA_2000} (top).

For solar PV, the following codes were considered to define eligible terrain:
1, 2, 3, 4, 5, 6, 7, 8, 9, 10, 11, 12, 13, 14, 15, 16, 17, 18, 19, 20, 26, 31, 32.
The resulting eligible terrain for solar PV after combining the Natura 2000 layer and the CORINE layer are shown in Figure \ref{CORINE_NATURA_2000} (bottom).

It is worth noticing that the overall electricity consumption in Spain could be satisfied by installing wind and solar in 1-2\% of the territory.

\

\newpage

%% file: 99_app_q2q_solar.tex
\section{Q2Q transform for solar PV generation} \label{app_q2q_solar}

Section \ref{subsec_Q2Q} provides an analysis on the application of the Q2Q transform to the onshore wind power generation in Spain. This Appendix contain the results of replicating this analysis for the case of the solar PV generation.

\begin{figure}[!ht]  
\centering
\includegraphics[width=14cm]{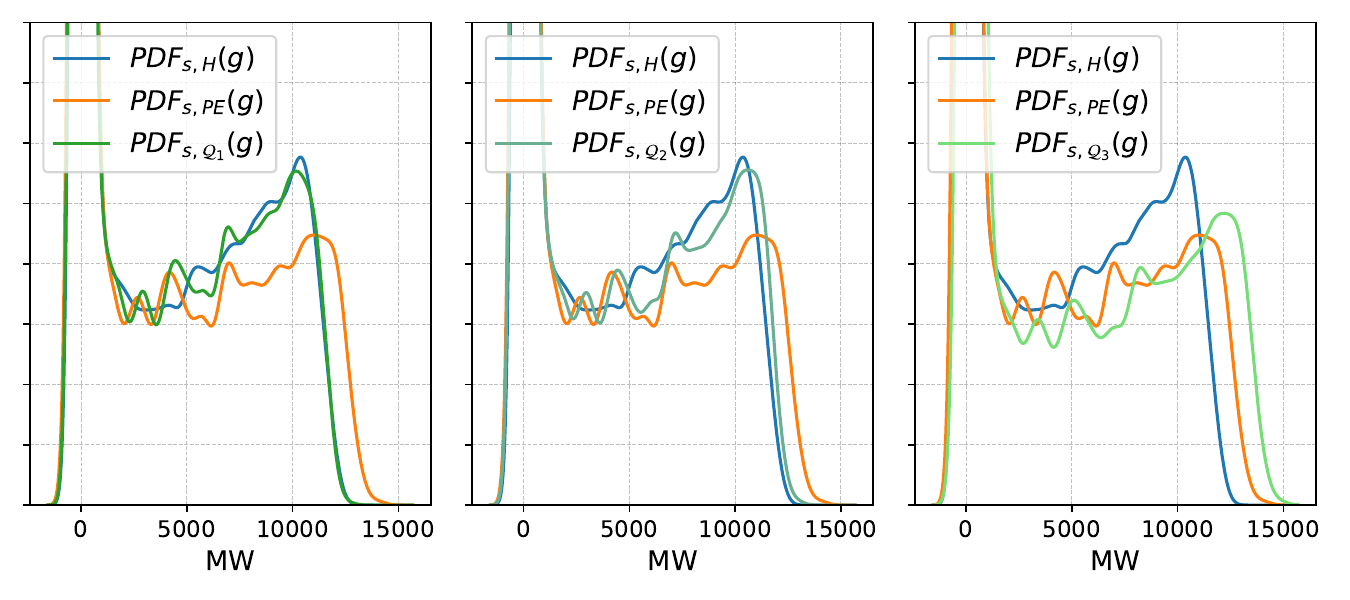}
\caption{PDFs of the solar PV generation time series in Spain in 2022. Subindices: $R$ stands for real data, $PE$ means modelled with PyPSA-Eur, and $\mathcal{Q}_i$ for $i=1,2,3$ refers to the Q2Q transform under normalisation scheme $i$.} \label{fig_q2q_solar_pdf}
\end{figure}

The obtained PDFs are illustrated in Figure \ref{fig_q2q_solar_pdf}. In this case, Q2Q transform under normalisation scheme 1 shows the best fitting, reducing the overestimation of high PV generation observed in the time series estimated with the PyPSA-Eur current methodology.

\begin{figure}[!ht]  
\centering
\includegraphics[width=12cm]{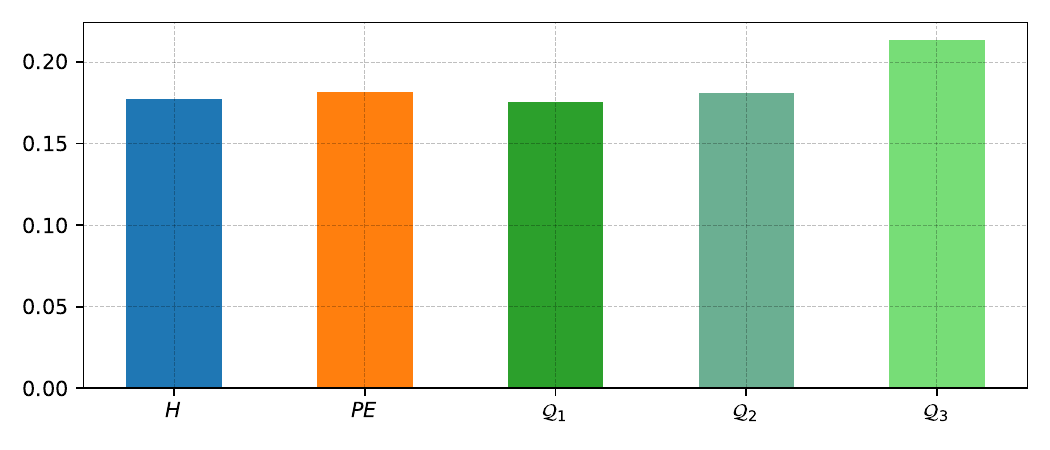}
\caption{CF of the wind power generation time series in Spain in 2022. Subindices: $R$ stands for real data, $PE$ means modelled with PyPSA-Eur, and $\mathcal{Q}_i$ for $i=1,2,3$ refers to the Q2Q transform under normalisation scheme $i$.} \label{fig_q2q_solar_CF}
\end{figure}

Figure \ref{fig_q2q_solar_CF} shows the annual capacity factor for each case. Normalisation scheme 3 results particularly bad.

\begin{figure}[!ht]  
\centering
\includegraphics[width=14cm]{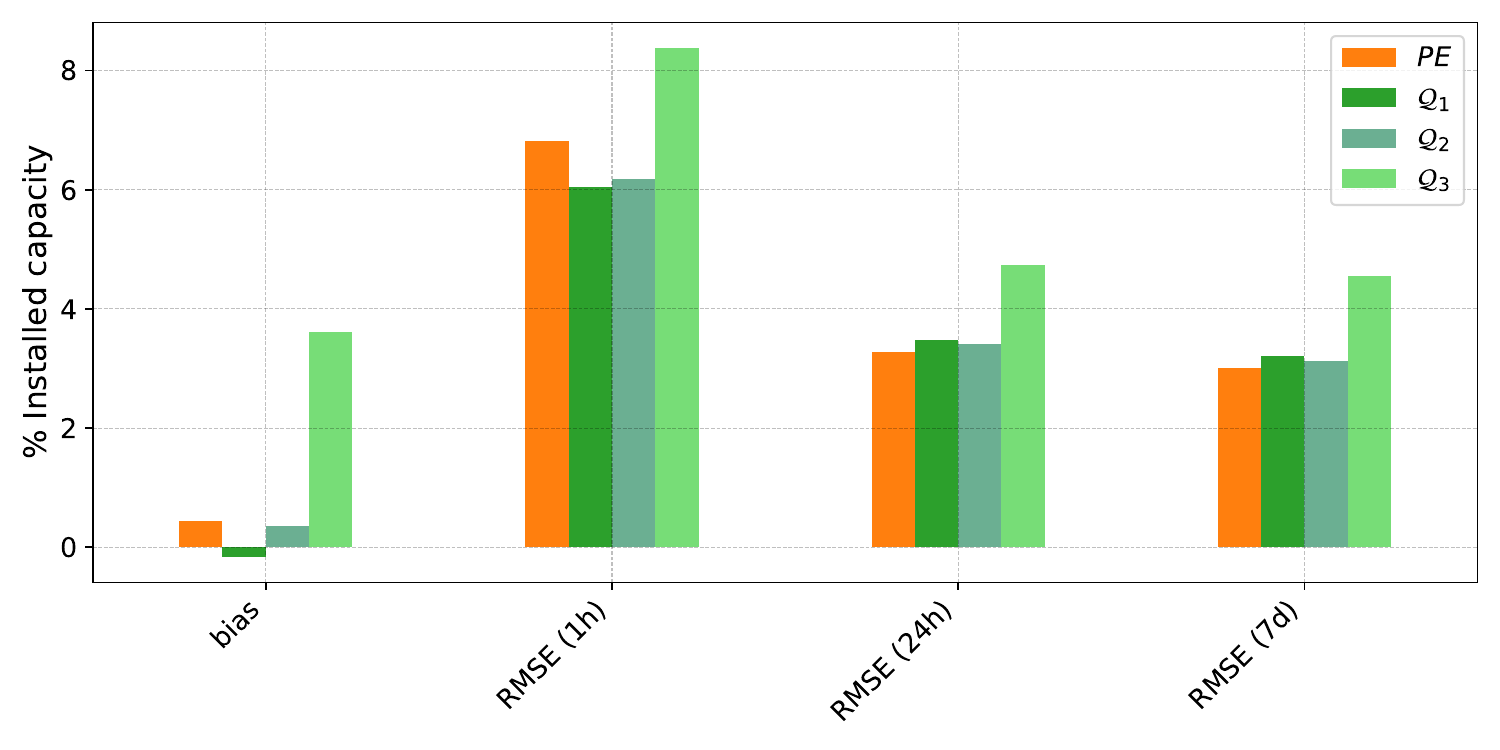}
\caption{Bias and RMSE of the simulated wind power generation time series in Spain in 2022. Subindices: $PE$ means modelled with PyPSA-Eur, and $\mathcal{Q}_i$ for $i=1,2,3$ refers to the Q2Q transform under normalisation scheme $i$.} \label{fig_q2q_solar_RMSE}
\end{figure}

Concerning the accuracy in reproducing the real solar PV generation over time, Figure \ref{fig_q2q_solar_RMSE} shows the bias and the RMSE for the different simulated cases. Normalisation schemes 1 and 2 provide the best performance in terms of RMSE, but lower bias is obtained with scheme 1.

%% file: 99_app_load.tex
\section{Supplementary information concerning spatial distribution of electricity demand in Spain} \label{app_load}

\begin{figure}[!ht]  
\centering
\includegraphics[width=5.3cm]{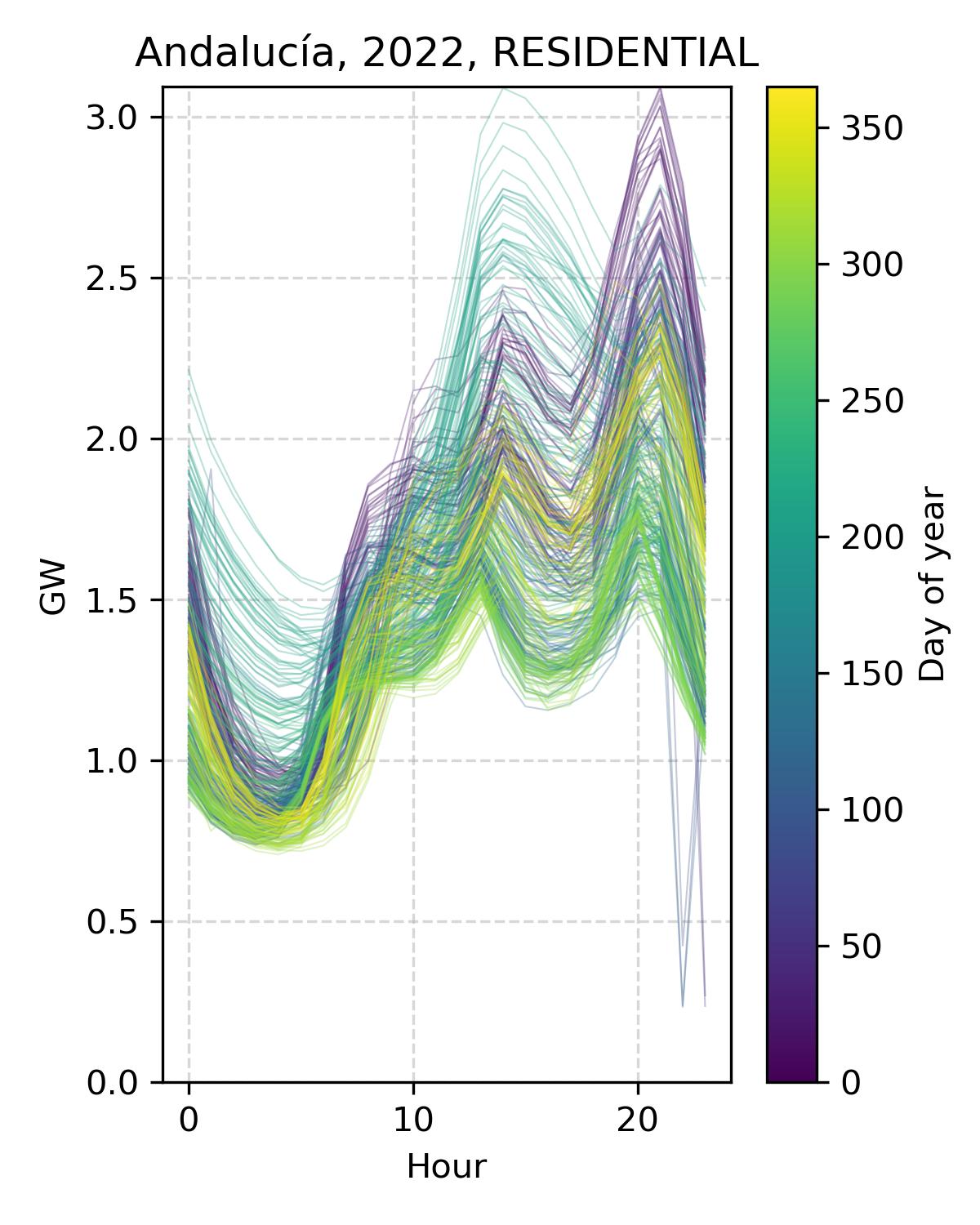}
\includegraphics[width=5.3cm]{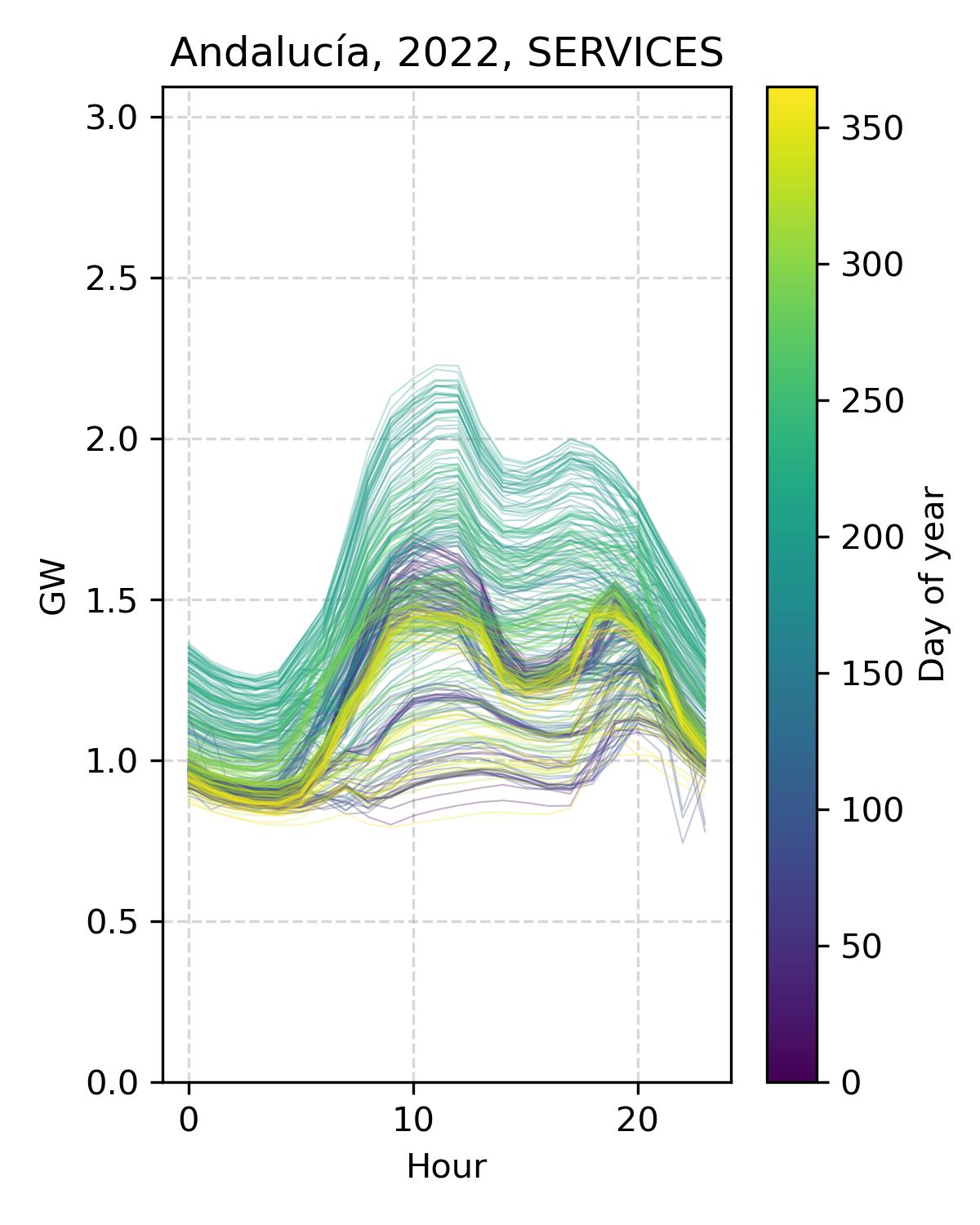}
\includegraphics[width=5.3cm]{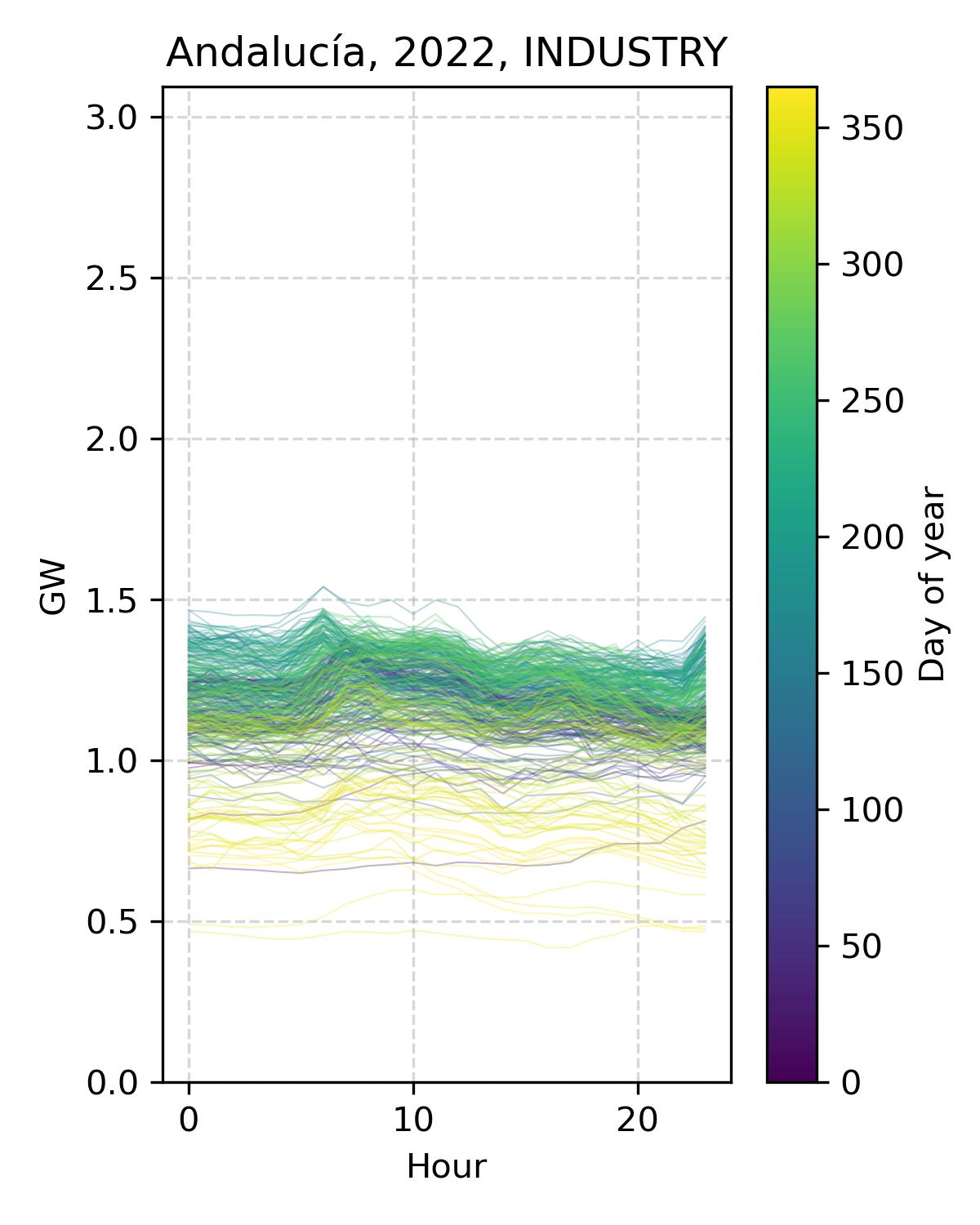}
\caption{Hourly electricity demand in Andalucía (NUTS 2 region in South Spain) according to economic sectors: residential, services and industry.} \label{fig_load_andalucia}
\end{figure}

\begin{figure}[!ht]  
\centering
\includegraphics[width=5.3cm]{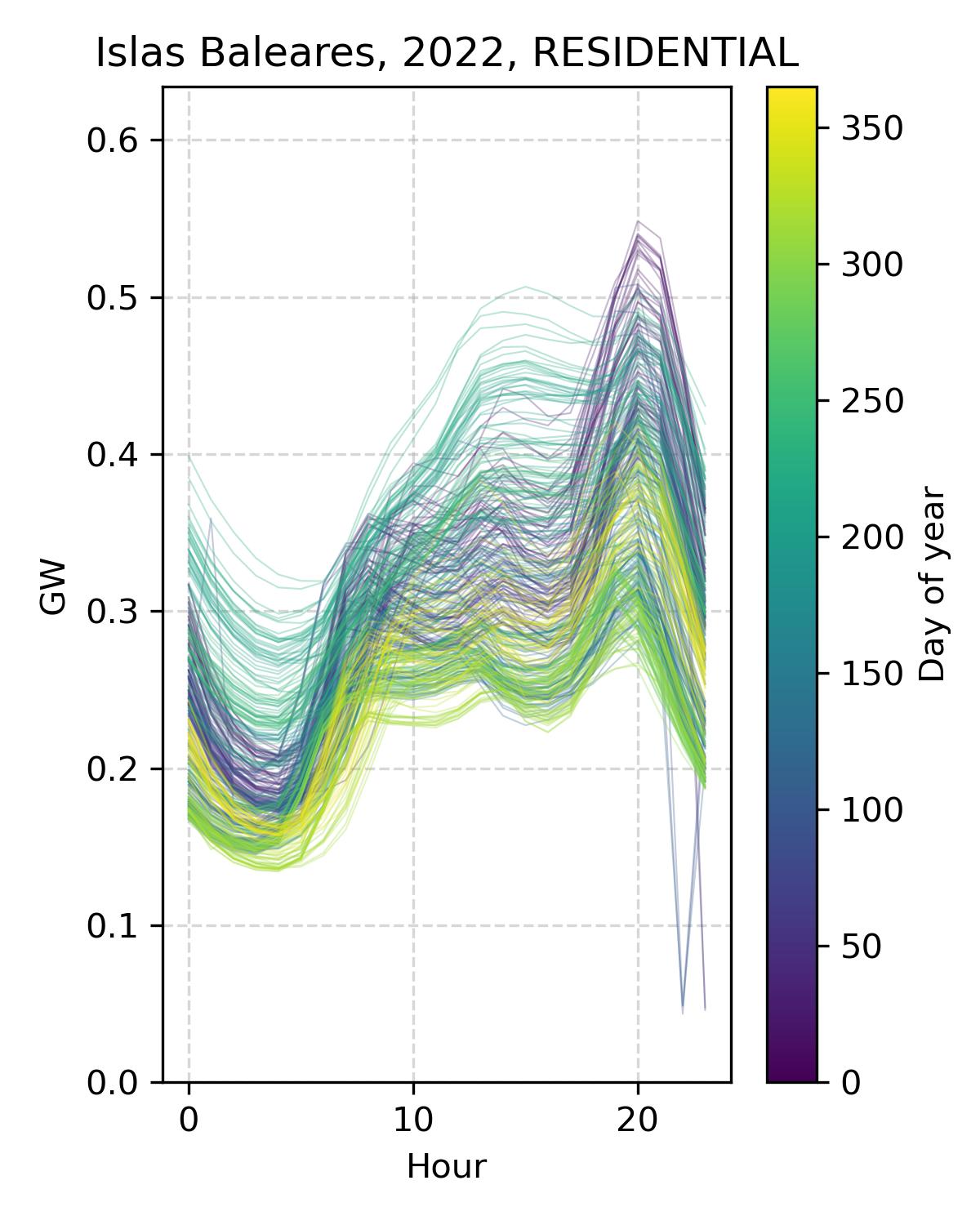}
\includegraphics[width=5.3cm]{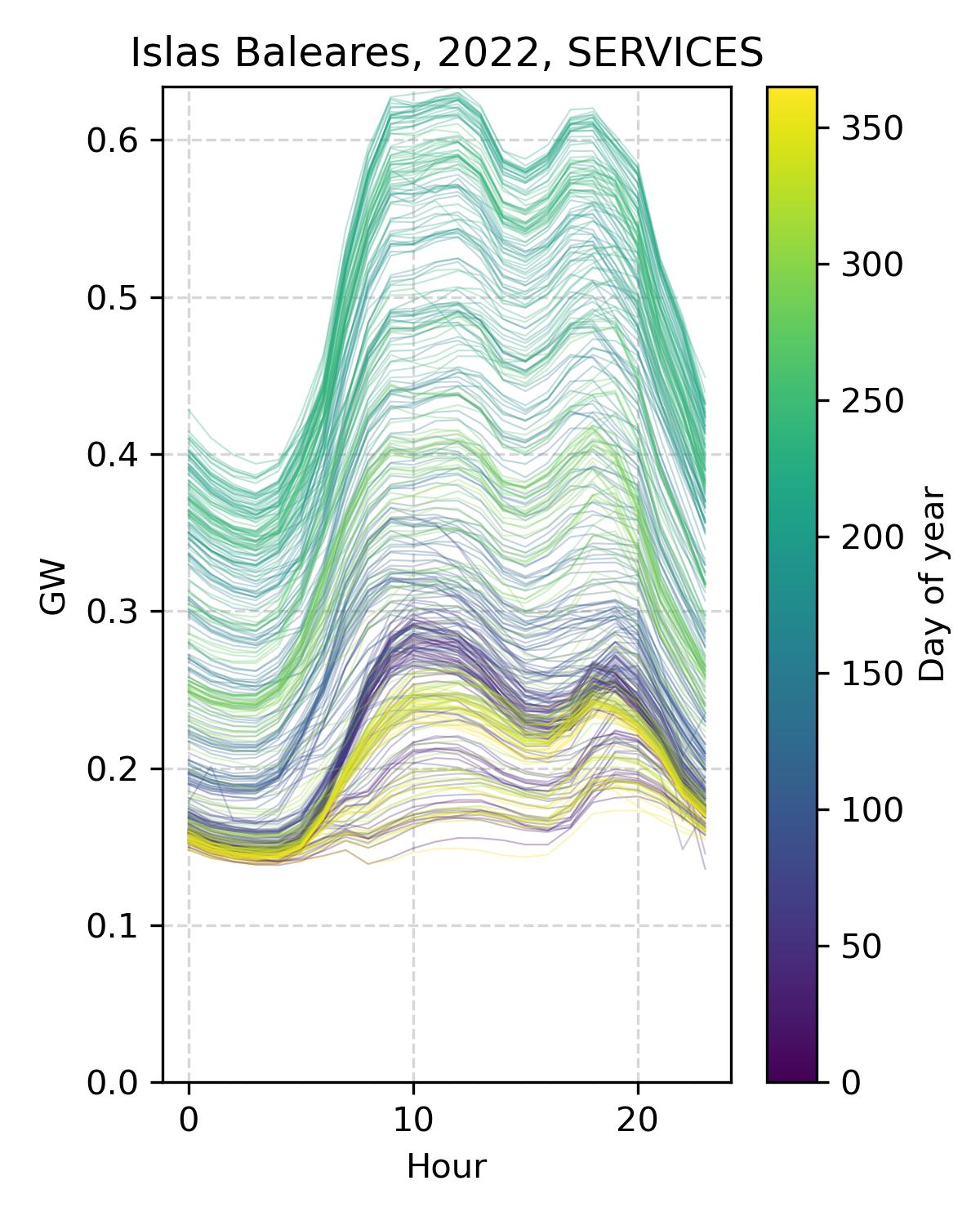}
\includegraphics[width=5.3cm]{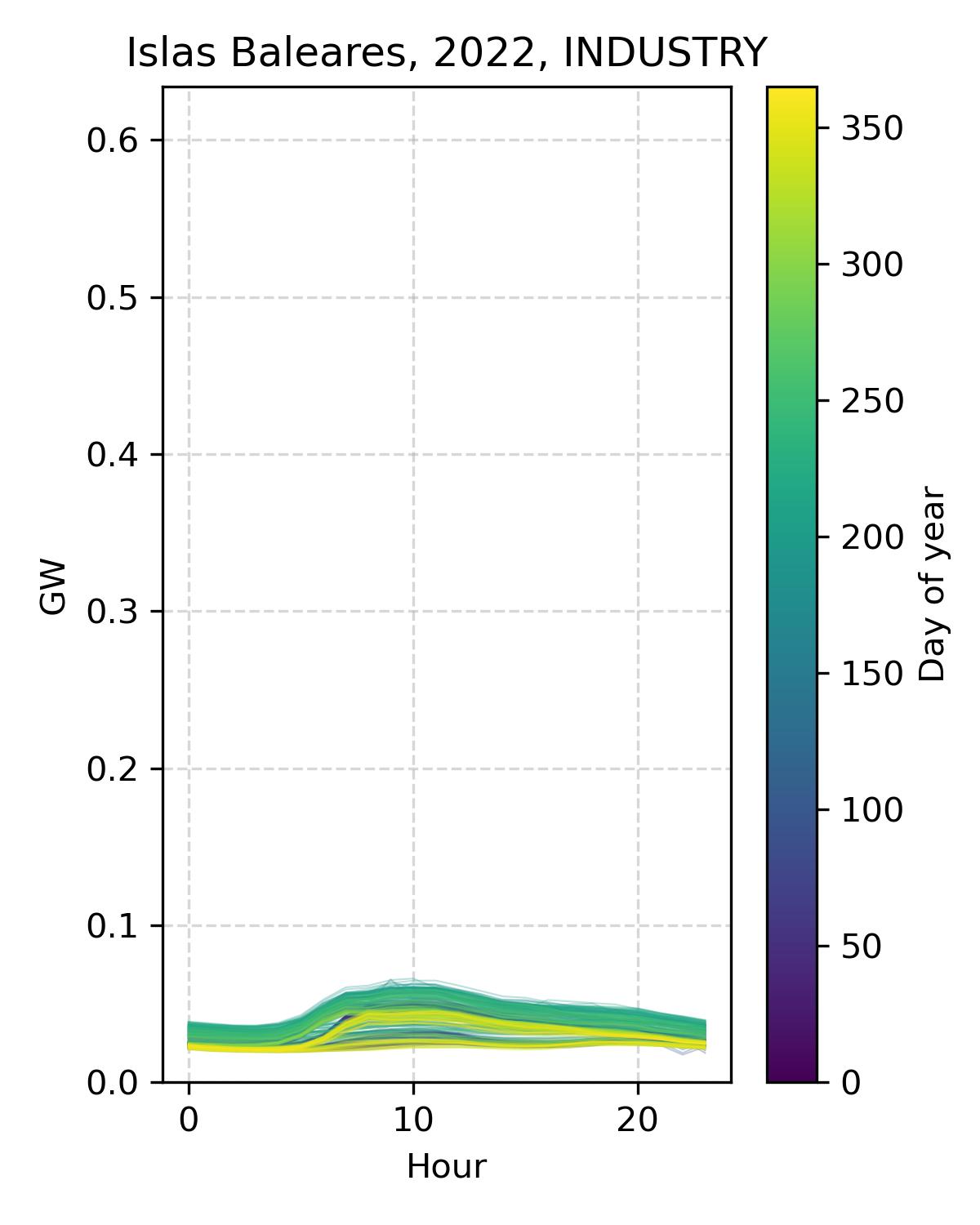}
\caption{Hourly electricity demand in Balearic Islands (NUTS 2 region in  East Spain) according to economic sectors: residential, services and industry.} \label{fig_load_baleares}
\end{figure}

\begin{figure}[!ht]  
\centering
\includegraphics[width=5.3cm]{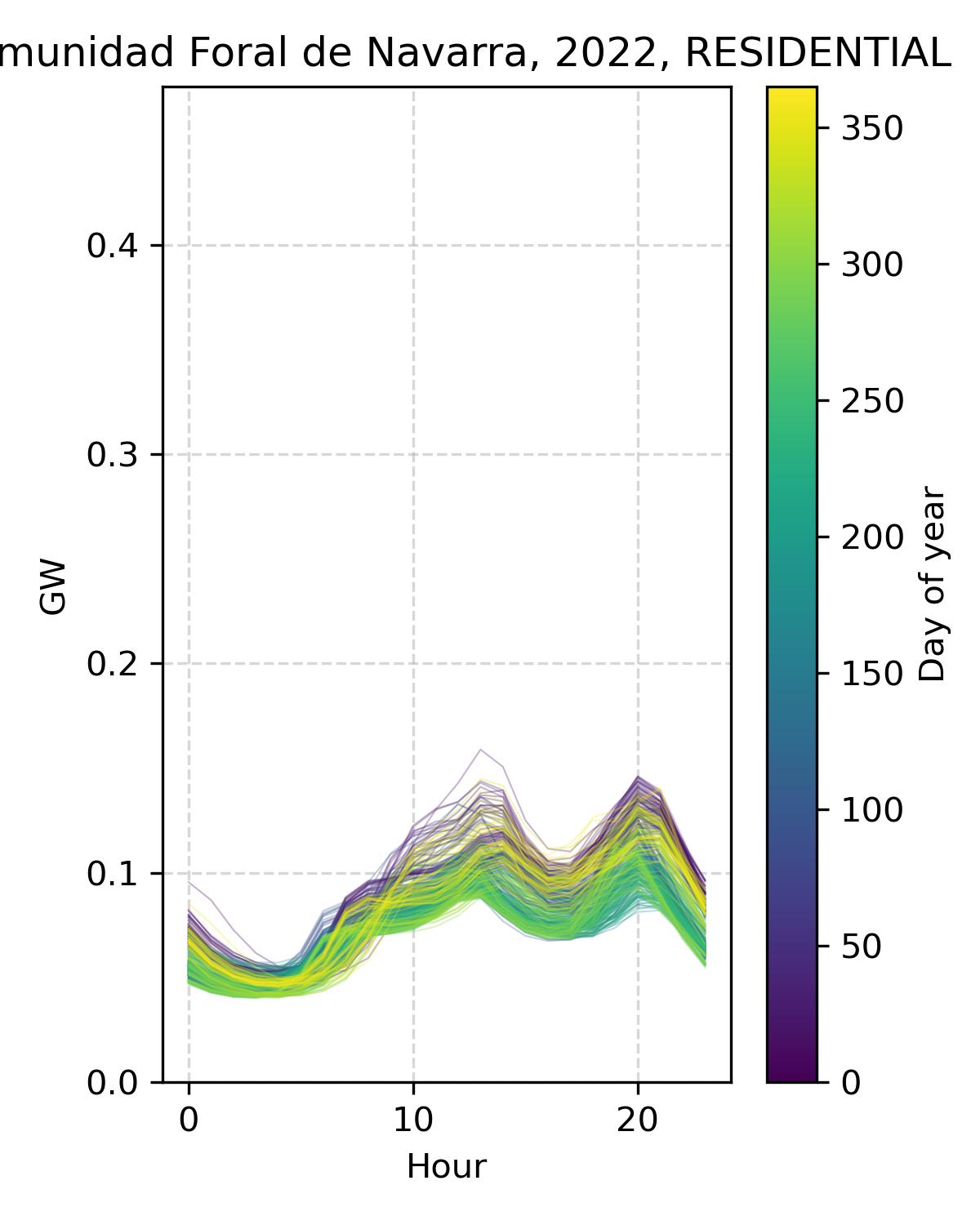}
\includegraphics[width=5.3cm]{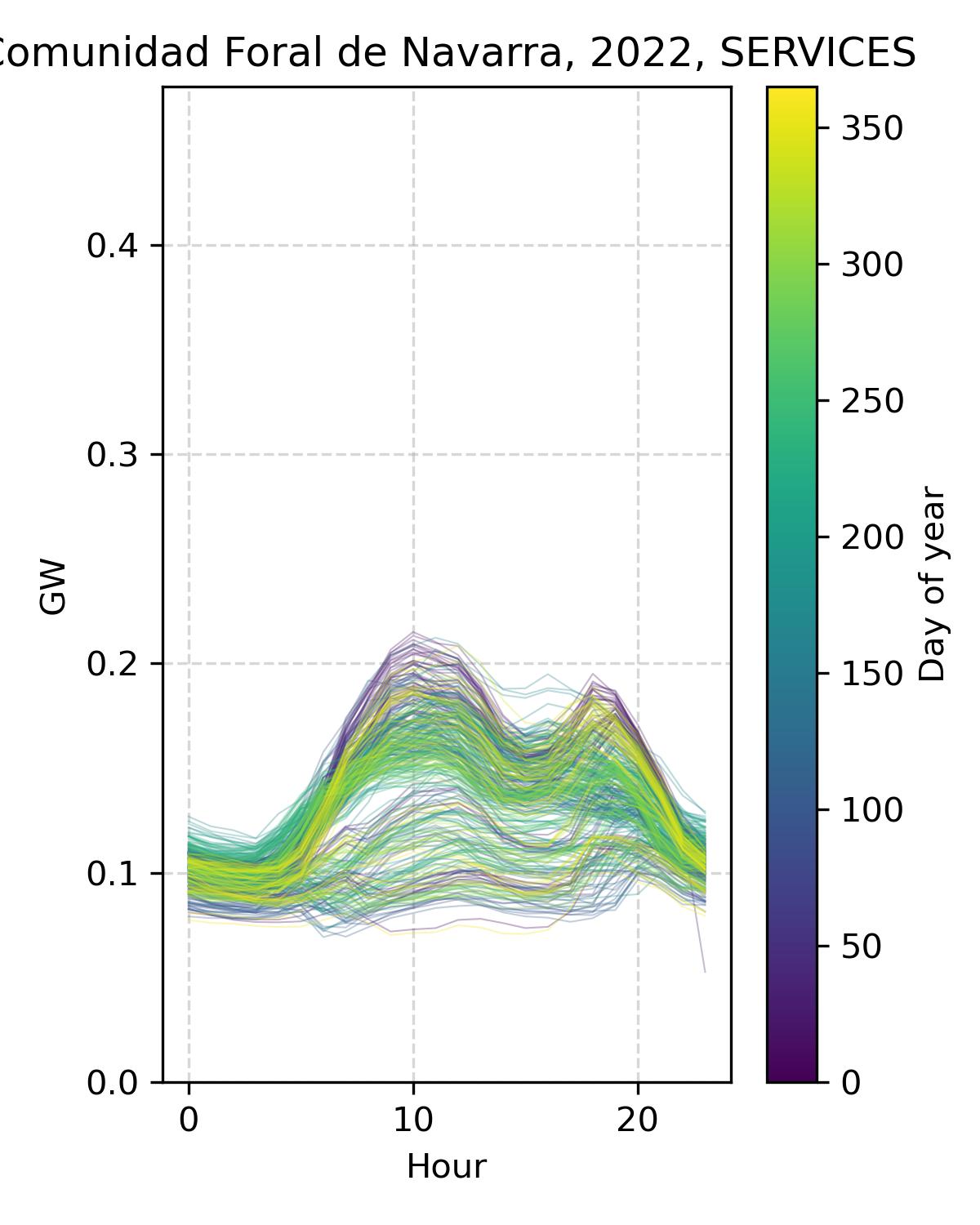}
\includegraphics[width=5.3cm]{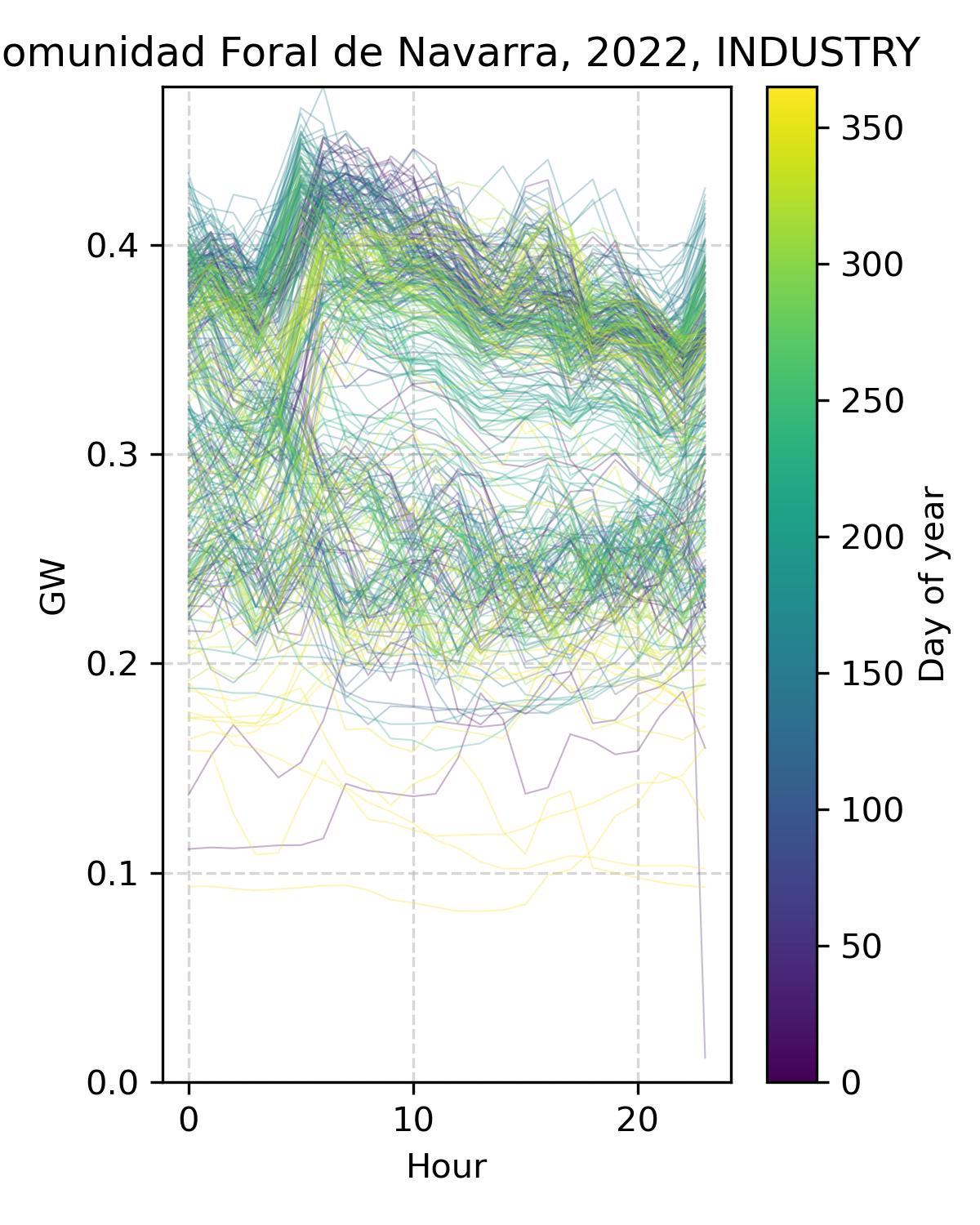}
\caption{Hourly electricity demand in Navarra (NUTS 2 region in  North Spain) according to economic sectors: residential, services and industry.} \label{fig_load_navarra}
\end{figure}

%% file: 99_app_PNIEC_NUTS3.tex
\section{Optimal mix for the 2030 CO2 target by Spanish provinces (NUTS 3 level) } \label{app_PNIEC_NUTS3}

\begin{figure}[!ht]  
\centering
\includegraphics[width=10cm]{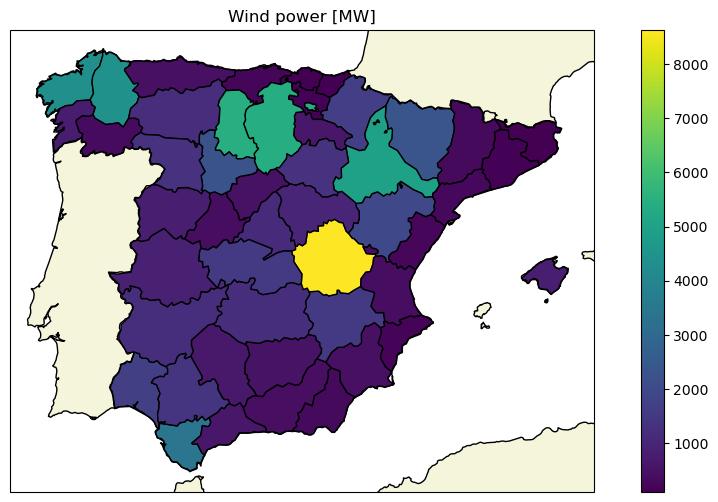}
\caption{Optimal onshore wind power capacity for the 2030 CO2 target, by Spanish provinces (NUTS 3 level).} \label{fig_PNIEC_NUTS3_onwind}
\end{figure}

\begin{figure}[!ht]  
\centering
\includegraphics[width=10cm]{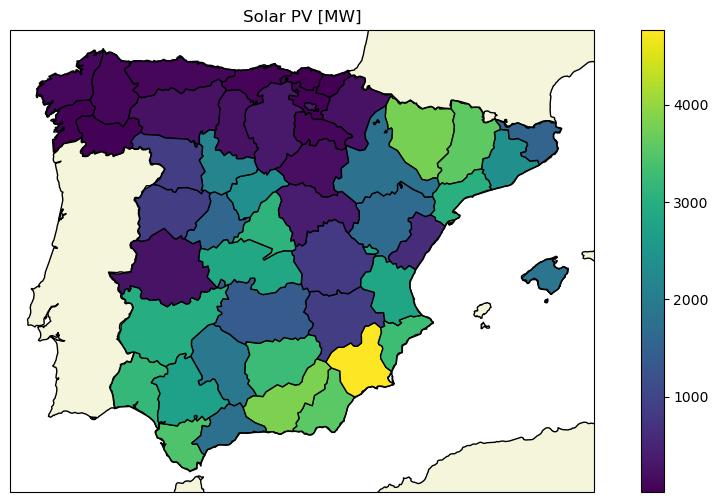}
\caption{Optimal solar PV capacity for the 2030 CO2 target, by Spanish provinces (NUTS 3 level).} \label{fig_PNIEC_NUTS3_solar}
\end{figure}

\begin{figure}[!ht]  
\centering
\includegraphics[width=10cm]{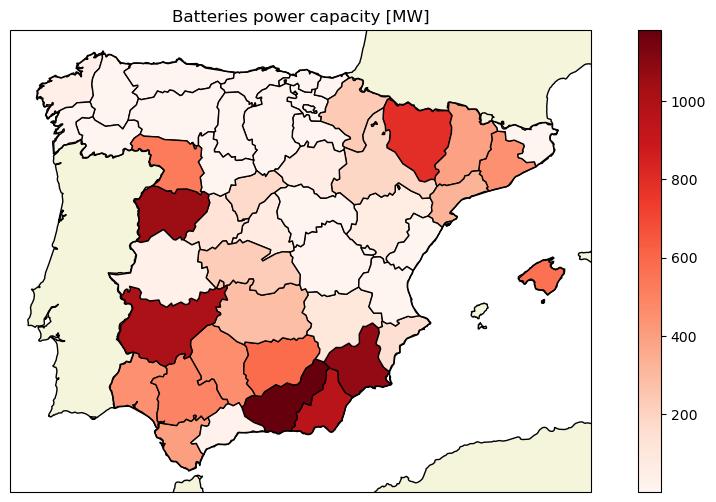}
\caption{Optimal batteries power capacity for the 2030 CO2 target, by Spanish provinces (NUTS 3 level).} \label{fig_PNIEC_NUTS3_batteries_power}
\end{figure}

\begin{figure}[!ht]  
\centering
\includegraphics[width=10cm]{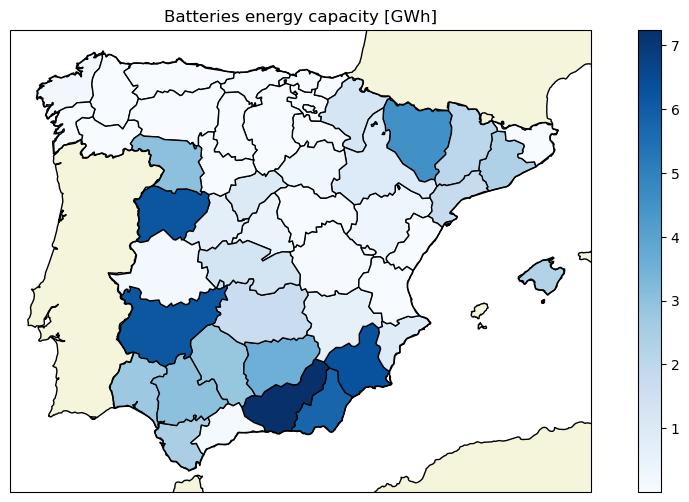}
\caption{Optimal batteries energy capacity for the 2030 CO2 target, by Spanish provinces (NUTS 3 level).} \label{fig_PNIEC_NUTS3_batteries_energy}
\end{figure}